\documentclass[twocolumn]{aastex62}
\usepackage{natbib,amsmath}
\usepackage{hyperref} \usepackage{float} 
\usepackage{afterpage}
\usepackage{subfigure}
\usepackage{xspace}
\usepackage{rotating}

\newcommand{\um}{\textmu m\xspace}

\providecommand{\e}[1]{\ensuremath{\times 10^{#1}}}

\begin{document}
\title{Retrievals on NIRCam transmission and emission spectra of HD 189733b with PLATON 6, a GPU code for the JWST era}

\correspondingauthor{Michael Zhang}
\email{mzzhang2014@gmail.com}

\author[0000-0002-0659-1783]{Michael Zhang}
\affil{Department of Astronomy \& Astrophysics, University of Chicago, Chicago, IL 60637}

\author[0000-0003-0062-1168]{Kimberly Paragas}
\affiliation{Division of Geological and Planetary Sciences, California Institute of Technology}

\author[0000-0003-4733-6532]{Jacob L.\ Bean}
\affil{Department of Astronomy \& Astrophysics, University of Chicago, Chicago, IL 60637}

\author{Joseph Yeung}
\affil{Department of Astronomy \& Astrophysics, University of Chicago, Chicago, IL 60637}

\author[0000-0003-1728-8269]{Yayaati Chachan}
\affil{Department of Physics and Trottier Space Institute, McGill University, 3600 rue University, H3A 2T8 Montreal QC, Canada}
\affil{Trottier Institute for Research on Exoplanets (iREx), Universit\'e de Montr\'eal, Canada}

\author[0000-0002-8963-8056]{Thomas P. Greene}
\affil{Space Science and Astrobiology Division, NASA Ames Research Center}

\author[0000-0003-2279-4131]{Jonathan Lunine}
\affil{Cornell University}

\author[0000-0001-5727-4094]{Drake Deming}
\affil{University of Maryland}

\begin{abstract}
We present the 2.4--5.0 \um JWST/NIRCam emission spectrum of HD 189733b, along with an independent re-reduction of the previously published transmission spectrum at the same wavelengths.  We use an upgraded version of PLanetary Atmospheric Tool for Observer Noobs (PLATON) to retrieve atmospheric parameters from both geometries.  In transit, we obtain [M/H]=$0.53_{-0.12}^{+0.13}$ and C/O=$0.41_{-0.12}^{+0.13}$, assuming a power-law haze and equilibrium chemistry with methane depletion.  In eclipse, we obtain [M/H]=$0.68_{-0.11}^{+0.15}$ and C/O=$0.43_{-0.05}^{+0.06}$, assuming a clear atmosphere and equilibrium chemistry without methane depletion.  These results are consistent with each other, and with a re-run of our previously published joint retrieval of HST and Spitzer transmission and emission spectra.  Accounting for methane depletion decreases the C/O ratio by 0.14/0.04 (transmission/emission), but changing the limb cloud parameterization does not affect the C/O ratio by more than 0.06.  We detect H$_2$O, CO$_2$, CO, and H$_2$S in both the NIRCam transmission and emission spectra, find that methane is depleted on the terminator, and confirm with VULCAN that photochemistry is a potential cause of this depletion.  We also find tentative (1.8$\sigma$) evidence of a dayside thermal inversion at millibar pressures.  Finally, we take this opportunity to introduce a new version of PLATON.  PLATON 6 supports GPU computation, speeding up the code up to 10x.  It also supports free retrievals using both volume mixing ratio and centered-log ratio priors; emission from planetary surfaces of different compositions; updated opacities at improved resolution; and Pareto smoothed importance sampling leave-one-out cross validation (PSIS-LOO).
\end{abstract}

\section{Introduction}
HD 189733b \citep{bouchy_2005}, the closest transiting hot Jupiter, has been extensively studied by both observers and theorists since its discovery in 2005.  Owing to its early discovery date and exceptional observational favorability, it has accumulated extensive transmission and emission spectroscopy data from multiple space telescopes, including HST/STIS \citep{sing_2011,evans_2013}, HST/WFC3 \citep{gibson_2012,mccullough_2014,crouzet_2014}, Spitzer in all four IRAC bands \citep{tinetti_2007,beaulieu_2008,agol_2010,desert_2011,morello_2014,knutson_2007,charbonneau_2008,agol_2010,knutson_2012}, and Spitzer/MIPS at 24 $\mu$m \citep{knutson_2009}.  The first ever phase curve observation was of HD 189733b \citep{knutson_2007}, as was the first ever detection of an exoplanet transit in X-rays \citep{poppenhaeger_2013,king_2019}.

On the modelling side, extensive work has been done on the planet's non-equilibrium chemistry (e.g. \citealt{moses_2013,tsai_2021}), cloud patterns (e.g. \citealt{lee_2015,lines_2018}), atmospheric escape (e.g. \citealt{lampon_2021}), and 3D circulation (e.g. \citealt{flowers_2019}), among other aspects.  The large signal sizes permit precise tests of model predictions.  For example, the hotspot offset was predicted before any phase curve had been observed of any exoplanet.  Repeat Spitzer eclipse observations constrained the planet's emission variability to less than 1.6\% at 4--5 \um.  The abundance of transit and eclipse observations is ideal for retrieving atmospheric properties.  In \cite{zhang_2020}, we performed a comprehensive retrieval on all published HST and Spitzer transmission and emission spectra, obtaining a [M/H] of $1.08 \pm 0.23$ and a C/O ratio of $0.66_{-0.09}^{+0.05}$--among the smallest uncertainties reported for an exoplanet in the pre-JWST era.  The retrieval found a terminator dominated by nanometer-sized grains, consistent with the prediction by \cite{ohno_2020} that photochemical hazes can generate super-Rayleigh scattering slopes at equilibrium temperatures of 1000-1500 K.

Since 2022, JWST has been revolutionizing exoplanet atmospheric observations.  JWST has expanded both the precision of transmission and emission spectra and their wavelength coverage manyfold, leading to a flood of new discoveries.  HD 189733b has been observed by JWST in transit or eclipse ten times by four separate programs (GTO 1185 by PI Greene, GTO 1274 by PI Lunine, GO 1633 by PI Deming, GO 2001 by PI Min, GO 2021 by PIs Kilpatrick and Kataria), using NIRCam, MIRI/LRS, and MIRI/MRS.  The NIRCam transmission spectrum from GO 1633 has already been published \citep{fu_2024}, revealing features from H$_2$O, CO$_2$, CO, and H$_2$S, but not from CH$_4$, which they find is heavily depleted from its equilibrium abundance.

A large number of exoplanet atmospheres will need to be observed in order to find patterns predicted by formation models--for example involving trapping of solids at ice lines \citep{oberg_2011}--to be tested.  Leaving aside the many processes other than condensation that determine disk composition, such as vertical temperature gradients, thermal and photochemical processes, and migration, it is logical to ask how well JWST data can measure metallicity and C/O ratio in the first place.  Certainly high SNR, high quality JWST data is no guarantee of precise constraints, or even of any meaningful constraints, as the Early Release Science MIRI/LRS phase curve of WASP-43b showed \citep{bell_2024}.  However, not every wavelength range is as information-poor as MIRI/LRS, and since our retrieval on pre-JWST HD 189733b transmission and emission data already yielded precise constraints, one might wonder how much more constraining JWST data would be.

To find out, we re-reduce the NIRCam 2.4--5.0 \um transit data with our own pipeline, Simple Planetary Atmosphere Reduction Tool for Anyone (SPARTA), and retrieve atmospheric properties with PLanetary Atmospheric Tool for Observer Noobs (PLATON).  We also reduce the new NIRCam 2.4--5.0 \um eclipse data from GTO 1185 and 1274, run a retrieval on the resulting spectrum, and compare it to the transmission retrieval, our pre-JWST (HST + Spitzer) retrieval, and retrieval results in the literature.  Just as \cite{zhang_2020} both introduced a major PLATON update and performed a retrieval on HD 189733b to demonstrate its new capabilities, this paper will both introduce the updates that make PLATON suitable for the JWST era, and use the new code to retrieve on JWST data of HD 189733b.

Section \ref{sec:observations} introduces the NIRCam observations, Section \ref{sec:reduction} discusses the data reduction, Section \ref{sec:platon} introduces the new features of PLATON, Section \ref{sec:retrievals} uses the new PLATON to retrieve on the NIRCam transmission and emission spectra, and Section \ref{sec:discussion} compares the results to literature values while commenting on the robustness of atmospheric retrievals.

\section{Observations}
\label{sec:observations}
\begin{table*}[htbp]
    \caption{NIRCam transit and eclipse observations of HD 189733b}
    \centering
    \setlength{\tabcolsep}{6pt}
    \begin{tabular}{c c c c c c}
        \hline
        Type & Filter & Date & Read Pattern & Groups & Dur. (h)\\
	\hline
        Transit & F322W2 & 2022-08-29 & BRIGHT1 & 3 & 5.9\\
        Transit & F444W  & 2022-08-25 & BRIGHT1 & 4 & 6.0\\
        Eclipse & F322W2 & 2022-08-26 & BRIGHT2 & 2 & 5.6\\
        Eclipse & F444W  & 2022-09-28 & BRIGHT2 & 4 & 5.6\\
        \hline
    \end{tabular}
    \label{table:observations}
\end{table*}

The NIRCam Grism Time Series observations we analyze consist of one transit and one eclipse in each of the F322W2 (2.4--3.95 \um) and F444W (3.90--5.00 \um) bandpasses, for a total of four visits.  Note the small 0.05 \um overlap between our conservatively defined bandpasses.  The observations come from three programs: GO 1633 (PI: Drake Deming), which observed both transits; GTO 1274 (PI: Jonathan Lunine), which observed the F322W2 eclipse; and GTO 1185 (PI: Thomas Greene), which observed the F444W eclipse.  The transits have been previously published in \cite{fu_2024}.

All four visits use the SUBGRISM64 subarray, the GRISMR pupil, and the NRCALONG detector, as is standard for grism time series observations of bright stars.  Their durations, read patterns, number of groups/integration, and dates of observation are listed in Table \ref{table:observations}.  The transit/eclipse duration is $1.84 \pm 0.04$ h \citep{addison_2019}, so all observations have ample baseline on both sides.  All observations were taken within a month of each other, and all except the F444W eclipse were taken within 4 days.  The gap between the two transits--4.4 days, or 2 orbital periods--is 37\% of the stellar rotation period \citep{henry_2008}, meaning that the side of the star facing the Earth, and therefore the pattern of spots and faculae, is expected to be very different.  All four visits include short-wavelength photometry at 2.1 \um, but we did not analyze this data because previous work has found it to be anomalous.  For example, \cite{fu_2024} excluded it because the SW transit depth is inconsistent with the spectroscopic transit depths measured by the grism, and \cite{xue_2024} excluded it because of the extremely high scatter ($\sim10\times$ photon noise).

All the JWST/NIRCam data used in this paper can be found in MAST:  \dataset[10.17909/d77d-tt16]{http://dx.doi.org/10.17909/d77d-tt16}.

\section{Data reduction}
\label{sec:reduction}
We reduced the NIRCam data using Simple Planetary Atmosphere Reduction Tool for Anyone (SPARTA), first introduced in \cite{kempton_2023} to analyze MIRI/LRS data of GJ 1214b, and first used to analyze NIRCam data of HD 149026b \citep{bean_2023} and 209458b \citep{xue_2024}.  SPARTA is fully independent of every other pipeline, including the official JWST pipeline.  We briefly describe the pipeline here, and refer the reader to these papers for more details.

We start with the uncalibrated files, and apply the standard calibration procedures: superbias subtraction, reference pixel subtraction, non-linearity correction, dark subtraction, multiplication by the gain (to convert DN to electrons), and ramp fitting.  The median residuals from the ramp fitting step across all integrations (an array with dimensions $N_{\rm grp} \times N_{\rm rows} \times N_{\rm cols}$) are computed and subtracted from the raw uncalibrated data, and the ramp fitting step is repeated, this time with $>14\sigma$ outliers being rejected.  This two-step ramp fitting procedure removes the uncorrected non-linearity (due mostly to the brighter-fatter effect--\citealt{plazas_2018}), which enables better cosmic ray detection and rejection.

After ramp fitting, we remove the background.  We first remove 1/f noise as much as possible by subtracting the median of the unilluminated columns for each row (columns 1894--2044 for F322W2, 4--600 for F444W).  This does not perfectly remove 1/f noise because the noise varies rapidly along each row, and because each row is spanned by four amplifiers while the unilluminated pixels cover only one (1.16 for F444W).  We then remove wavelength-dependent background by subtracting the median of the background rows (defined as 4-11 and 57-64) for each column.

After background fitting, we create a median image by taking the pixel-wise median of all integrations.  We use this image as a template to determine the trace x and y positions for each integration.  In addition, we use the image as the profile for optimal extraction, which we perform with a window half-height of 5 pixels while iteratively rejecting $>5\sigma$ outliers.  The per-integration spectra thus obtained are combined with the per-integration x and y positions into a single file, which serves as the input data file for the MCMC fits.

The MCMC fits in question begin with a fit to the white light curve for all four visits.  The systematics model consists of an exponential ramp and a linear function of x, y, and time:

\begin{align}
    S = F_* (1 + A\exp{(-t/\tau)} + c_y y + c_x x + m(t-\overline{t})),
\end{align}

The free transit parameters are the transit time, transit depth, $a/R_*$, b, and quadratic limb darkening parameters.  For the F322W2 transit, we mask the starspot crossing (defined as phases -0.000519 to +0.00695).  After fitting the two white light curves, we find that the time of transit they imply are different by $7.5 \pm 1.1$ s when propagated to the same epoch.  This cannot be due to the period uncertainty, which contributes only 25 milliseconds.  We assume the discrepancy is due to correlated noise in NIRCam and average the two transit times to obtain a more accurate time, which we adopt for fitting the spectroscopic light curves.  We obtain fully consistent $a/R_*$ ($8.882 \pm 0.014$ and $8.886 \pm 0.015$ for F322W2 and F444), as well as fully consistent $b$ ($0.6660 \pm 0.0016$ and $0.6671 \pm 0.0019$).  We average the two values of $a/R_*$ and $b$ to obtain the values that we adopt for the spectroscopic light curves in both transit and eclipse.

\begin{figure*}[ht]
  \centering \subfigure {\includegraphics
    [width=0.48\textwidth]{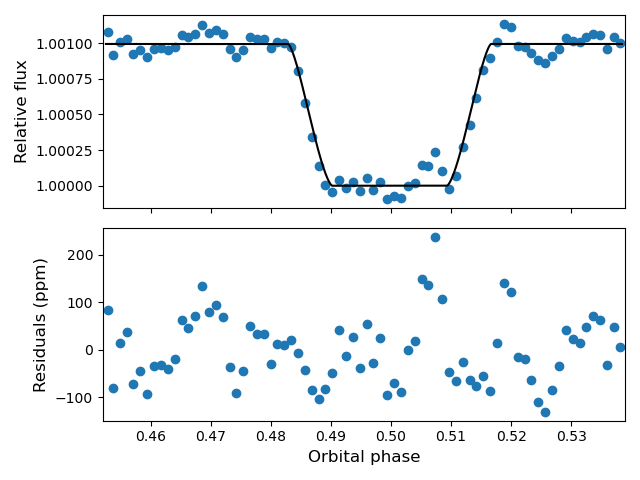}}\qquad\subfigure {\includegraphics
    [width=0.48\textwidth]{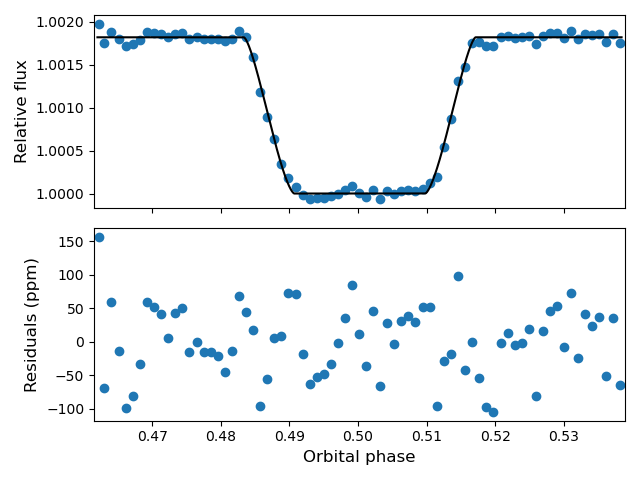}}
    \caption{Binned white light curve of the eclipse observation in the F322W2 filter (left) and the F444W filter (right), with the best fit systematics model divided out, the best fit eclipse model plotted on top, and the residuals at the bottom.  Note the significantly worse correlated noise in the bluer band.}
\label{fig:eclipse_white_lc}
\end{figure*}

\begin{figure}[ht]
  \includegraphics
    [width=0.48\textwidth]{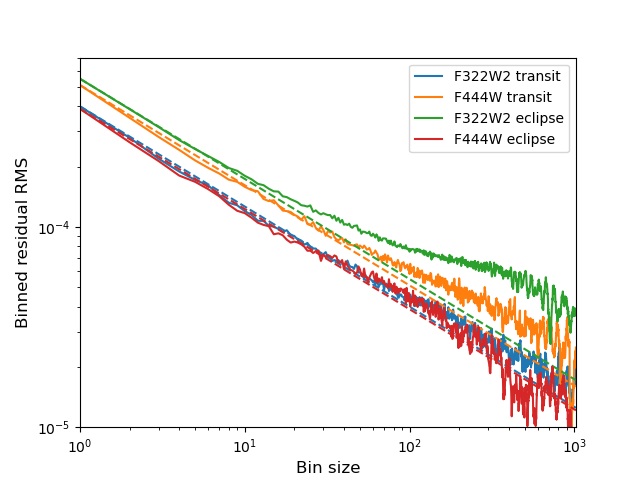}
    \caption{Allan deviance of the two transit and two eclipse white light curves.  The F322W2 eclipse, the only visit that uses two groups, has substantially higher RMS than all other visits at high bin sizes.  However, this visit uses a read pattern of BRIGHT2, so the 2 groups consist of 2 averaged reads each; whereas the F322W2 transit uses 3 groups with the BRIGHT1 pattern, for a total of only 3 reads.}
\label{fig:allan_deviances}
\end{figure}

The free eclipse parameters are the eclipse time and eclipse depth.  From the two white light curves, we infer eclipse time offsets of $-43 \pm 19$ s and $47 \pm 8$ s from phase 0.5, offsets which are discrepant at $4.3\sigma$.  It is not impossible that some part of this discrepancy is astrophysical--a negative $e\cos{\omega}$ combined with a wavelength-dependent hot spot offset could conceivably explain the discrepancy.  However, the hot spot offset measured by Spitzer is larger at 3.6 \um than at 4.5 \um \cite{knutson_2012}.  In addition, when we allow the eclipse time offset to be free while fitting the spectroscopic light curves, the offsets do not agree in the 4.0 \um overlap region: they continue to be negative for F322W2 and positive for F444W.  For these reasons, we assume that the discrepancy is due to correlated noise and set the offset to 0 while fitting the spectroscopic light curves.

The two eclipse white light curves are shown in Figure \ref{fig:eclipse_white_lc}, and Figure \ref{fig:allan_deviances} shows the corresponding Allan deviance plots for both the transit and eclipse visits.  As can be seen, the F322W2 eclipse is plagued by substantial correlated noise, including a 150 ppm bump right before the end of eclipse.  This shows up in the Allan deviance plot as a flattening of the curve at $\sim$50 ppm.  The F444W eclipse light curve, on the other hand, is remarkably clean: not only is there no sign of correlated noise in the white light residuals, the RMS comes very close to decreasing as the square root of the bin size, even at a bin size of 1000.  This is all the more remarkable because a bin size of 1000 corresponds to 0.85 h for the redder band, compared to only 0.47 h for the bluer band.  In the retrievals, we will account for the lower reliability of the F322W2 eclipse depths by fitting for an instrumental offset that applies to the blue band only, while assuming that the F444W depths do not need adjustment.  Figure \ref{fig:allan_deviances} also shows that the F322W2 transit exhibits much lower RMS than the F322W2 eclipse.  Perhaps not coincidentally, the F322W2 eclipse is the only visit of the four to use two groups.  We recommend caution in interpreting NIRCam data taken with only two groups.

After obtaining the transit parameters ($R_p/R_*$, $a/R_*$, $b$, transit time, eclipse time) from the white light fits, we fix them to the best fit values while fitting the spectroscopic light curves, which were binned to 50 nm.  The MCMC fits to the spectroscopic light curves have the same systematics model with the same priors as the fits to the white light curves.  For the spectroscopic transit light curves, the limb darkening coefficients are handled the same way as in \cite{fu_2024}: we fix the first quadratic coefficient to the values given by ExoTiC-LD using the 3D stagger grid \citep{magic_2015}, while the second quadratic coefficient is free and has a flat prior.

Unlike in \cite{fu_2024}, we do not attempt to correct for the transit light source effect or nightside pollution.  Their Extended Data Figure 6 shows that while nightside pollution is a minor effect, stellar inhomogeneity correction brings the F444W transit depths down by $\sim$120 ppm relative to the F3222W2 depths.  Instead of attempting these corrections by assuming a nightside spectrum, a stellar spot/facula coverage fraction, and a spot/facula spectrum, we will fit for an offset between the two transit spectra during our retrievals, which removes the vast majority of the impact of nightside pollution and time-varying stellar inhomogeneity.

\begin{figure}[ht]
  \includegraphics
    [width=0.5\textwidth]{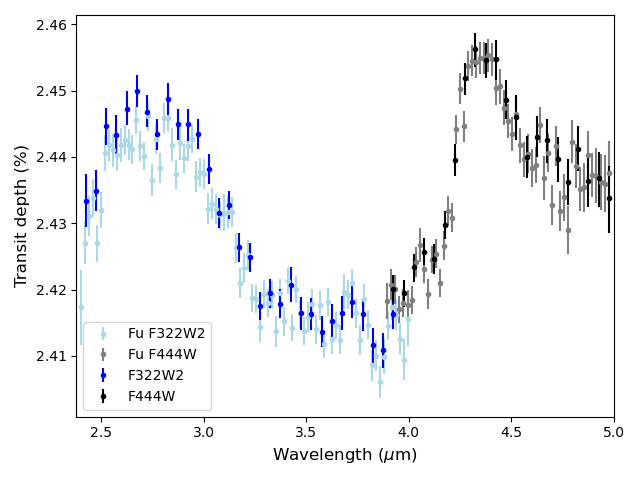}
    \caption{Comparison of the fiducial transmission spectrum adopted by \citep{fu_2024} to that adopted by the present paper (dark colors).}
\label{fig:transit_spec_comp}
\end{figure}

\begin{figure}[ht]
    \includegraphics[width=0.48\textwidth]{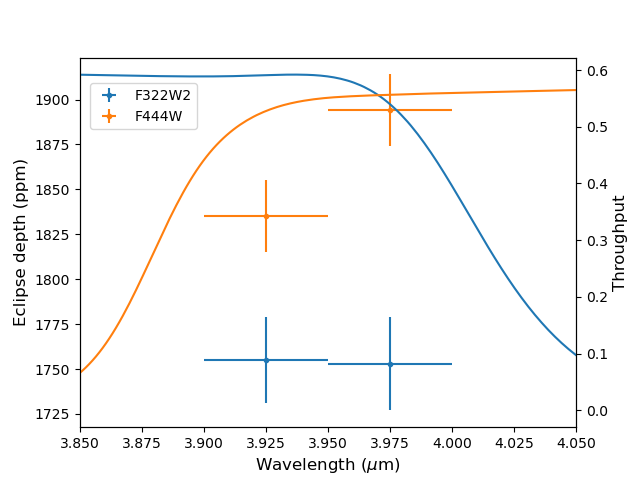}
    \caption{Comparison of the eclipse depths in the overlap region between the two filters.  The curves indicate the system throughput for the two filters.  Note the large discrepancy for the reddest bin, and the sharp dropoff of the F322W2 throughput at that wavelength.}
    \label{fig:discrepant_points}
\end{figure}

Figure \ref{fig:transit_spec_comp} compares the fiducial transmission spectrum in \cite{fu_2024} to the one we obtained.  Just as in \cite{fu_2024}, SPARTA and G.Fu's custom pipeline give consistent spectra, but the SPARTA spectrum is slightly higher at the shorter wavelengths (2.4--3.0 \um).  Our spectrum differs from SPARTA spectrum in \cite{fu_2024} in two main ways: we adopt a high bin size of 50 nm to mitigate the impact of non-line-by-line opacity resolution in the retrievals; and we correct a major bug in the wavelength solution.  The bug arose in STScI's NIRCam wavelength calibration, which was off by $\sim$30 nm in some regions of both the F322W2 and F444W bandpasses prior to a calibration update in September 2023 (private communication, JWST Help Desk).

We noticed one peculiarity in the spectroscopic eclipse depths: in the overlap region of 3.9--4.0 \um, the two visits do not agree with each other (see Figure \ref{fig:discrepant_points}).  While the discrepancy is only 2.5$\sigma$ in the 3.90--3.95 \um bin ($1755 \pm 24$ vs. $1836 \pm 20$ ppm in the bluer and redder filters respectively), it grows to 4.3$\sigma$ in the 3.95--4.00 \um bin ($1753 \pm 26$ ppm vs. $1894 \pm 20$ ppm).  Disturbingly, visual examination of the four spectroscopic light curves and their residuals reveals nothing anomalous.  Nevertheless, we believe that the F444W filter is more reliable for the 3.95--4.00 \um bin.  This bin spans the very edge of the F322W2 bandpass, where the spectral response decreases steeply with wavelength, while the spectral response is much flatter for the F444W filter.  The RMS of the residuals is also higher for the F322W2 light curve.  Finally, as we saw in Figure \ref{fig:allan_deviances}, the F322W2 visit in general is plagued by much higher correlated noise.  We therefore exclude the F322W2 3.95--4.00 \um bin from the emission--and out of caution, we exclude it from the transmission spectrum as well.

\subsection{Comparison with Spitzer/IRAC photometry}
Over more than a decade, Spitzer Space Telescope's IRAC instrument has measured the channel 1 (3.2--3.9 \um) and channel 2 (4.0--5.0 \um) transit and eclipse depths for hundreds of exoplanets.  The comprehensive uniform re-analysis of \cite{deming_2023} contains no fewer than 457 eclipses from 122 exoplanets.  HD 189733b was one of the most extensively observed planets, with seven published eclipses in channel 2 \citep{kilpatrick_2020} and 9 in channel 1 (all homogenously analyzed in \citealt{deming_2023}), five channel 1 transits \citep{ehrenreich_2007,beaulieu_2008,desert_2009,desert_2011,knutson_2012}, and two channel 2 transits \citep{desert_2009,knutson_2012}.  Given Spitzer's strong, percent-level intrapixel sensitivity variations, the reliability of Spitzer measurements has long been an open question.

To compare our depth measurements to Spitzer's IRAC channel 1 and channel 2 depth measurements, we take the weighted average of the JWST spectroscopic transit/eclipse depths to derive an IRAC-equivalent broadband depth.  At each wavelength, the weight is the product of the stellar spectrum (converted from an energy flux to a photon flux) and the corresponding IRAC channel's spectral response.  If we assume the spectroscopic depths are independent, we can calculate the error on the weighted average mathematically, obtaining 5--7 ppm for all four combinations of geometry and IRAC channel.  However, since the spectroscopic depths are by no means independent--the noise is highly correlated from wavelength to wavelength--such a calculation severely underestimates the true error.  We instead perform a MCMC fit to the white light curve with the wavelength boundary adjusted to the IRAC channel's bandpass (3.20--3.87 \um for channel 1, 4.00--5.00 \um for channel 2), and adopt the error from the fit as a more realistic estimate.

In emission, we obtain an IRAC-equivalent broadband depth of $1302 \pm 16$ ppm for channel 1 and $1792 \pm 16$ ppm for channel 2.  The weighted average depth of the Spitzer eclipses is $1432 \pm 34$ in channel 1 \citep{deming_2023} and $1849 \pm 28$ in channel 2 (calculated by D. Deming based on the catalog in \citealt{deming_2023}).  Using the same seven channel 2 eclipses, \cite{kilpatrick_2020} report a depth of $1827 \pm 22$ ppm.  While the redder band is consistent between the two telescopes (to 1.8$\sigma$ and 1.3$\sigma$ for the former and latter analyses), the NIRCam IRAC-equivalent depth for channel 1 is 3.5$\sigma$ lower than the Spitzer measurement.  The discrepancy expands to 4.5$\sigma$ if one accounts for the -40 ppm offset favored by our retrieval (Section \ref{sec:retrievals}) for F322W2 eclipse depths.  The much larger discrepancy between JWST and Spitzer in channel 1 supports the long-standing conventional wisdom that channel 1 is less reliable for exoplanet observations.  This is likely due to its much larger intrapixel sensitivity variations--according to Spitzer documentation, it is 8.1\% across the central pixel for channel 1, and only 2.1\% for channel 2 \footnote{\url{https://irsa.ipac.caltech.edu/data/SPITZER/docs/irac/calibrationfiles/pixelphase/}}.  Another possibility is that Spitzer's channel 1 eclipse depth is closer to the truth: Figure \ref{fig:allan_deviances} shows that the F322W2 eclipse exhibits significant correlated noise while the F322W2 transit is much cleaner, possibly because the former was taken with only two groups while the latter was taken with three.  (As a caveat, the former's two groups consist of four reads because of the BRIGHT2 read pattern, while the latter's three groups consist of only three reads because of the BRIGHT1 read pattern.)

In transmission, we obtain an IRAC-equivalent broadband depth of $24,179_{-96}^{+110} \pm 90$ ppm in channel 1 and $24,402_{-110}^{+89}$ ppm in channel 2.  The much larger error bars compared to emission are largely due to the degeneracy between the transit depth and the limb darkening parameters.  Another factor hampering the inter-telescope comparison is that we are not aware of any homogeneous analysis of all the Spitzer transits using state-of-the-art methods.  We instead compare to \cite{pont_2013}, which averaged three of the five channel 1 measurements \citep{desert_2009,desert_2011,knutson_2012} and both channel 2 measurements \citep{desert_2009,knutson_2012} to obtain $24,047 \pm 84$ ppm and $24,155 \pm 109$ ppm respectively.  Although these are consistent with our NIRCam results to 2$\sigma$, this comparison is not as revealing as the eclipse depth comparison because variability in stellar flux, spot/facula crossings, and the uncertainty in the limb darkening parameters all impact the broadband transit depth. 
 Combined, they can easily explain discrepancies of the order of 100 ppm.

Previous eclipse depth comparisons between JWST and Spitzer for WASP-121b \citep{morello_2023} and WASP-77Ab \citep{august_2023} reported depths consistent at 2$\sigma$ for both bands.  However, neither paper took into account the different spectral responses of Spitzer and the JWST instrument.  \cite{morello_2023} compared NRS1 to IRAC channel 1 and NRS2 to IRAC channel 2, but the bandpasses are not exactly the same.  The uncertainties on the Spitzer eclipse depths are 80--90 ppm for WASP-121b and 60--80 ppm for WASP-77Ab, far above the uncertainties on the combined HD 189733b eclipses, making these comparisons less stringent than ours.  Comparisons between JWST and Spitzer were also performed for HD 140926b \cite{bean_2023}, finding 2$\sigma$ consistency with the two channel 2 Spitzer measurements and one of the two channel 1 Spitzer measurements.  On the other hand, the channel 1 depth from \cite{zhang_2017} is inconsistent with the JWST data at 4.4$\sigma$.  Perhaps not coincidentally, just as with our HD 189733b data, the statistical error bars for the channel 1 HD 149026b data are very small (20--30 ppm).

\section{PLATON 6}
\label{sec:platon}
In \cite{zhang_2020}, we presented our last major update to PLATON, which, among other improvements, added support for emission spectroscopy and updated opacities.  In the four years since 2020, JWST has launched and begun collecting exquisite spectra of exoplanet atmospheres.  The unprecedented precision of these measurements, which open up rocky planets to observational analysis, require model improvements in many areas.  For example, higher resolution opacities are necessary for gas giants in order for the model error induced by the limited resolution to be smaller than observational error.  This, in turn, means that the code needs to be faster in order for retrievals to finish in a reasonable time.  For rocky planets--super-Earths and smaller--the atmosphere is unlikely to be hydrogen/helium dominated, necessitating free retrievals where the background gas does not need to be specified a-priori.  These rocky planets may also have thin atmospheres or no atmosphere at all, making the surface emission important to calculate.  We discuss these and other updates in turn.  As with previous versions of PLATON, the new version is available on GitHub.\footnote{\url{https://github.com/ideasrule/platon}}

\subsection{Opacity update and GPU support}
Earlier PLATON releases downloaded R=1k opacities by default, although R=10k opacities were also available for manual download.  Due to the much higher precision of JWST data, PLATON now defaults to R=20k opacities.  The user can adopt even higher resolutions (higher than R=100k for $\lambda < 10$ \um) by downloading the appropriate opacities from the DACE opacity database\footnote{\url{https://dace.unige.ch/opacityDatabase/}}, converting the cross sections to absorption coefficients (in m$^{-1}$), and interpolating them to PLATON's temperature, pressure, and wavelength grid.  We provide 0.51--12 \um R=80k opacities online \footnote{\url{https://astro.uchicago.edu/~mz/absorption.html}}, where the molecular opacities are interpolated from DACE files and the alkali opacities are calculated using a combination of \cite{allard_2016,allard_2019} and the NIST database.  As with the previous release, we computed most of the default opacities with \texttt{ExoCross} \citep{yurchenko_2018} using the latest line lists.  The exceptions are Na, K, CH$_4$ and VO.  The alkalis have very strong non-Lorentzian line wings, which we compute using code from \cite{allard_2016,allard_2019}; while CH$_4$ and VO opacities were computed with HELIOS-K \citep{grimm_2021} because of the enormous number of transitions.  We tabulate the line lists we use to generate all opacities in Table \ref{table:line_lists}, and indicate whether each has been updated since \cite{zhang_2020}.  In addition to updating the line lists, we added opacities for several species (FeH and several metals), and expanded the range of the opacity files in wavelength and temperature to 0.2 $\mu$m and 4000 K respectively.  These changes are meant to improve PLATON's ability to model ultra-hot Jupiters, especially those with HST/WFC3 UVIS data in the 200--300 nm range.

The drastic increase in opacity resolution leads to a drastic increase in the number of computations required for each forward model.  To keep run time low, we use cupy to add GPU support.  cupy is intended to be a drop-in replacement for numpy.  Since PLATON was designed to use numpy array operations as much as possible for optimization purposes, in theory, changing all instances of numpy to cupy should be enough to convert PLATON to a GPU code.  In practice, cupy does not have all the functionality of numpy, and operations that are fast on the CPU may be unacceptably slow on the GPU, requiring us to make minor code changes.  With these changes, PLATON now runs on a GPU when cupy is installed, falling back on the CPU if cupy is not found.  Forward models are $\sim10$x faster with R=10k opacities, with even greater speedups at higher resolutions.  In \cite{zhang_2020}, our combined retrieval of the transit and eclipse spectrum of HD 189733b, spanning 0.32 \um to 25 \um, took 3 weeks on a Core i9 9900k (the top-end consumer CPU of the time).  With the new version of PLATON and a Core i9 13900k (the top-end consumer CPU of today) combined with the RTX 4090 (the top-tier gaming GPU), the same retrieval takes 2 days, despite a doubling of the opacity resolution.

\begin{table}[htbp]
    \centering
    \setlength{\tabcolsep}{6pt}
    \begin{tabular}{c c c c}
        \hline
        Species & Source & Updated? \\
	\hline
        C$_2$H$_2$ & ExoMol--aCeTY & New$^1$\\
        CH$_4$ & ExoMol--MM & Yes$^2$\\
        CO & ExoMol--Li2015 & No\\
        CO$_2$ & ExoMol--UCL-4000 & Yes$^3$\\
        H$_2$S & ExoMol--AYT2 & No\\
        H$_2$O & ExoMol--POKAZATEL & No \\
        HCN & ExoMol--Harris & No\\
        NH3 & ExoMol--CoYuTe & No\\
        NO & ExoMol--XABC & Yes$^4$\\
        NO2 & HITRAN 2020 & Yes$^5$\\
        OH & ExoMol--MoLLIST & No\\
        O2 & HITRAN 2020 & Yes$^5$\\
        O3 & HITRAN 2020 & Yes$^5$\\
        OCS & ExoMol--OYT8 & Yes$^6$\\
        PH3 & ExoMol--SAlTY & No\\
        SiH & ExoMol--SiGHTLY & No\\
        SiO & ExoMol--SiOUVenIR & Yes$^7$\\
        SO2 & ExoMol--ExoAmes & No\\
        TiO & ExoMol--ToTo & No\\
        VO & ExoMol--HyVO & Yes$^8$\\
        FeH & ExoMol--MoLLIST & New$^9$\\
        Na & Allard+19 & New$^{10}$\\
        K  & Allard+16 & New$^{11}$\\
        Ca & NIST & New$^{12}$\\
        Ti & NIST & New$^{12}$\\
        Fe & NIST & New$^{12}$\\
        Ni & NIST & New$^{12}$\\
        \hline
    \end{tabular}
    \caption{Line lists used to generate opacity files.  References for new or updated line lists: 1) \citealt{chubb_2020}, \citealt{yurchenko_2024}, 3) \citealt{yurchenko_2020}, 4) \citealt{qu_2021}, 5) \citealt{gordon_2022}, 6) \citealt{owens_2024}, 7) \citealt{yurchenko_2022}, 8) \citealt{bowesman_2024}, 9) \citealt{bernath_2020}, 10) \citealt{allard_2019}, 11)\citealt{allard_2016}, 12) NIST Atomic Spectra Database}
    \label{table:line_lists}
\end{table}

\subsection{Free retrieval with VMR and CLR priors}
The next major addition to PLATON is free retrieval capability.  Because PLATON was written with simplicity in mind, it has always been easy to hack in this capability, and many users have done so (e.g. \citealt{ih_2021,august_2023}).  With PLATON 6, we make this capability official and support two different priors: log volume mixing ratio (log VMR) and centered-log-ratio (CLR).  In the former, all gases in the atmosphere except one (the ``filler gas'') have a log-uniform prior on their mixing ratio, while the filler gas' mixing ratio is calculated so that the sum is 1.  The implicit prior on the filler gas is strongly biased toward higher mixing ratios.  To see this, imagine that we have two non-filler gases with log-uniform mixing ratio priors between $10^{-12}$ and 1.  With these priors, the mixing ratio of both gases is lower than 0.01 with high probability; therefore, the sum of the mixing ratios is lower than 0.02 with high probability.  The mixing ratio of the filler gas would then be greater than 0.98 with high probability.  For atmospheres where the filler gas is known to dominate the composition (e.g. hot Jupiters), this prior is acceptable, even desirable; for atmospheres where the dominant gas is unknown, such as super-Earths, we would like a prior that treats all gases equally.

Centered-log-ratio (CLR) priors are designed to treat all gases equally.  This prior formulation has a long history in geology (e.g. \citealt{aitchison_1982}), and was first applied to exoplanet atmospheres by \cite{benneke_2012}.  The CLR $\xi_i$ of gas i in a mixture of n gases is defined as:

\begin{align}
    \xi_i &= \ln{X_i} - \ln{g},
\end{align}

where    
\begin{align}
    g &= \exp{\Big(\frac{1}{n} \sum_{j=1}^{n} \ln{X_i}\Big)},
\end{align}

where $X_i$ is the mixing ratio of gas i.  We adopt uniform priors on $\xi_1,\xi_2,...,\xi_{n-1}$.  Since $\sum \xi_i = 0$, $\xi_n$ is calculated as negative the sum of the other CLRs.  Since $\sum X_i=g \sum e^{\xi_i} = 1$, we can calculate $g=1/\sum{\xi_i}$, after which we calculate individual mixing ratios as $X_i = g e^{\xi_i}$.  The remaining crucial step is to reject all samples where any $X_i$ is below a lower threshold, which we adopt as $10^{-12}$.

When using nested sampling, it is not possible to have a uniform prior with infinite bounds.  We therefore have to determine the minimum and maximum possible CLR.  The maximum CLR is achieved when all gases are at $X_{min}$, dragging $\ln{g}$ as negative as possible, while the remaining gas is at an abundance close to 1.  The equations above then give the CLR of the remaining gas, which is very close to $-\ln{g}$: $\xi_{\rm max} \approx -\frac{n-1}{n} \ln{X_{min}}$.  The minimum CLR is achieved when one gas is at $X_{min}$ and the remaining n-1 gases have abundances close to 1/(n-1), thus dragging $\ln{g}$ as high as possible.  The equations above give the CLR of the minimum-abundance gas: $\xi_{\rm min} \approx \frac{n-1}{n} (\ln{X_{min}} + \ln{(n-1)})$.  By adopting uniform priors between these two limits and setting the likelihood of samples where any gas ratio is less than $X_{min}$ to -inf, we guarantee two desirable properties: (1) all gases have the same prior, and (2) at low mixing ratio, the prior reduces to a uniform prior on log(VMR).

\subsection{Modelling surface emission}
Aside from adopting an appropriate prior, another important step in enabling rocky planet retrievals is modelling the surface flux, which will be visible if the atmosphere is thin or non-existent.  We model the surface flux, prior to absorption by the atmosphere, as a sum of reflection and emission:

\begin{align}
    F_{\rm surf} = (1-A_{g,\lambda}) \pi B_\lambda(\lambda, T_s) + A_{g,\lambda} F_{*,\lambda} (R_*/a)^2,
\end{align}

The wavelength-dependent albedos are inferred from hemispherical reflectance data of different rocky materials that we measured in the lab, which will be described in detail in Paragas et al. (in prep).  Surface emission functionality should be considered in beta until Paragas et al. (in prep) is accepted.


\subsection{Bayesian leave-one-out cross validation}
In the updated PLATON, every retrieval will report the expected log probability density for every data point (elpd$_{\rm LOO}$) using a Bayesian leave-one-out cross-validation method, namely Pareto Smoothed Importance Sampling (PSIS).  This metric was first used for exoplanet retrievals by \cite{welbanks_2023}, although it has a long history in other fields (e.g. \citealt{vehtari_2012}).  The most correct way to calculate elpd$_{\rm LOO}$ would be to leave each data point out in turn, re-run the retrieval, and see how well the left-out data point is predicted, but this would be prohibitively expensive.  PSIS is a method of estimating elpd$_{\rm LOO}$ solely from the log likelihood contributed by each data point during each sample of the nested sampling or MCMC run.  In PLATON, we use the Python implementation of PSIS-LOO by \cite{vehtari_2015} (\url{https://github.com/avehtari/PSIS}).

Intuitively, elpd$_{\rm LOO}$ is how well the retrieved model would predict a given data point if that point was left out of the retrieval.  It therefore reflects (but is not equivalent to) how much a given data point affects the retrieval: if a given point has high elpd, including it would generally not change the retrieval result very much, while if it has low elpd, it may substantially change the result once included.  However, it is possible that a point with low elpd may not change the result at all because it is such an outlier that no model comes close to predicting it.

One way to use elpd, shown in \cite{welbanks_2023}, is to compute the $\Delta$elpd between two retrievals, say a reference retrieval with CO$_2$ and an alternative one without.  elpd$_{\rm ref}$ - elpd$_{\rm no CO_2}$ would then be a measure of how much better each data point can be predicted from the other data points once CO$_2$ is included.  More loosely, points with a high $\Delta$elpd are the ones primarily driving the CO$_2$ detection.

\subsection{Miscellaneous changes}
Previous versions of PLATON generated their equilibrium abundance grids using GGchem \citep{woitke_2018}.  In this version of PLATON, we switch to FastChem \citep{kitzmann_2024} because of its speed and robustness.  Additionally, we make two slight changes to our definition of metallicity and C/O ratio: we update our solar abundances from \cite{asplund_2009} to \cite{asplund_2021}, and we scale both C and O abundances so that their sum, instead of the O abundance alone, respects the specified metallicity.

Previous versions of PLATON used \texttt{dynesty} \citep{speagle_2019} for nested sampling.  However, while running HD 189733b retrievals for this paper, we noticed that \texttt{dynesty} posteriors were sometimes substantially different from run to run, even when we used 1000 live points.  This problem has also been seen with another retrieval code, SCARLET (private communication, C. Piaulet).  We therefore added support for an alternative nested sampling package, pymultinest, a Python wrapper around MultiNest \citep{feroz_2009,feroz_2019}.  We previously avoided pymultinest because it was difficult to install, but the ubiquity of anaconda has made installation trivial: conda install -c conda-forge pymultinest.  In our experience, pymultinest is faster than \texttt{dynesty} when the same number of live points is used.  This is especially the case for our HD 189733b transmission spectrum retrievals, for which dynesty made several times more likelihood calls than pymultinest.  We leave detailed investigation of the performance and robustness of different nested sampling methods to future work.

Previous versions of PLATON used collisionally induced absorption (CIA) data inherited from Exo-Transmit \citep{kempton_2017}, which in turn was based on HITRAN.  HITRAN has CIA data for many molecule pairs, but the temperature range is extremely limited and extremely cold for all pairs except H$_2$-H$_2$ and H$_2$-He.  For example, H$_2$-CH$_4$ only has data in the 40--400 K range, and O$_2$-O$_2$ only has data at 293 K or 296 K for the 7450--14898 cm$^{-1}$ wavenumber range.  The previous PLATON CIA data was therefore based on very speculative extrapolation from very cold temperatures.  In this version of PLATON, we remove all molecule pairs from the CIA data except H$_2$--H$_2$, which has data from 200--3000 K, and H$_2$--He, which has data from 200-9900 K.

PLATON supports Mie scattering hazes with a constant user-specific complex refractive index, or with wavelength-dependent refractive indices corresponding to real materials.  In this version of PLATON, we expand the inventory of supported haze compositions from three (solid MgSiO$_3$, amorphous SiO$_2$, and TiO$_2$) to all condensates with data in LX-MIE \citep{kitzmann_2018}, except Fe$_2$SiO$_4$ and MgAl$_2$O$_4$.  These two were excluded because they do not have data down to 0.2 \um.

Finally, we add plotting scripts to PLATON to plot a variety of useful quantities, such as the contribution function, optical depths, the temperature/pressure profile, and abundance profiles.  It has always been possible for users to plot these quantities themselves, but the scripts decrease the learning curve and guide the user to the useful atmospheric parameters output by PLATON.

\section{Retrievals on HD 189733b}
\label{sec:retrievals}
Using our updated version of PLATON, we perform retrievals on the transmission and emission spectra of HD 189733b.  We perform 6 different retrievals: one on each of the visits individually, one on the combined transmission spectrum, and one on the combined emission spectrum.  We then compare the results to our comprehensive pre-JWST retrieval on this planet \citep{zhang_2020} (hereafter Z20), which incorporated all of the most up-to-date HST and Spitzer transmission and emission spectra.  The pre-JWST retrieval was re-run with the updated opacities and FastChem abundances of the newest version of PLATON.  All retrievals were run with pymultinest with 1000 live points and importance sampling enabled.

In our retrievals, we assume that the atmosphere is in chemical equilibrium, and parameterize it with log metallicity ([M/H]) and C/O ratio.  Equilibrium condensation is taken into account, so that no species is supersaturated, and the metallicity and C/O ratio include atoms in condensates.  The transit retrievals have as free parameters the planet radius, [M/H], C/O, limb temperature, offset between the two transits, aerosol parameters, and error inflation factor (a constant which multiplies all error bars).  We adopt parameterized Rayleigh scattering ($\sigma_\lambda = A(\lambda/\lambda_0)^{-p}$) plus an opaque cloud deck as our fiducial aerosol model, with three free parameters: log($A$), the log of the scattering cross section (relative to Rayeigh scattering at 3 \um); $p$, the spectral slope; and log cloud-top pressure of the opaque cloud deck.  All parameters were given uniform priors, with the following bounds: 1.11--1.13 $R_J$ for radius; -1--3 for [M/H]; 0.2--2 for C/O; 500--1300 K for limb temperature; -200--200 ppm for the offset; -1--10 for the log of the scattering factor; 3--12 for the scattering slope; -1--8 for the log cloudtop pressure (in Pa); and 1--2 for the error multiple.  The upper limit of 12 on the scattering slope is somewhat arbitrary, as the NIRCam transmission spectrum is consistent with slopes at least as high as 25.  However, since HST/STIS optical transmission spectra are consistent with a Rayleigh-like slope of 4 \citep{pont_2013,zhang_2020}, we considered a slope above 12 to be highly implausible.  Widening the prior does not change either [M/H] or C/O by more than $\sim$0.05.

The emission retrieval fixes the planet radius to that found by Z20.  The free parameters are [M/H], C/O, an offset for the F322W2 eclipse depths, an error inflation factor, and the same five temperature-pressure profile parameters from the parameterization of \cite{line_2013} that we fit for in Z20: $\log{\kappa_{th}}$, $\log{\gamma}$, $\log{\gamma_2}$, $\alpha$, $\beta$.  As in Z20, we assume a clear atmosphere because the hot day side is less conducive to cloud formation than the colder limb, and because the observed emission is inherently weighted toward the hotter and less cloudy portions of the day side.

\subsection{Metallicity and C/O: full equilibrium}

\begin{figure*}[ht]
  \centering \subfigure {\includegraphics
    [width=0.48\textwidth]{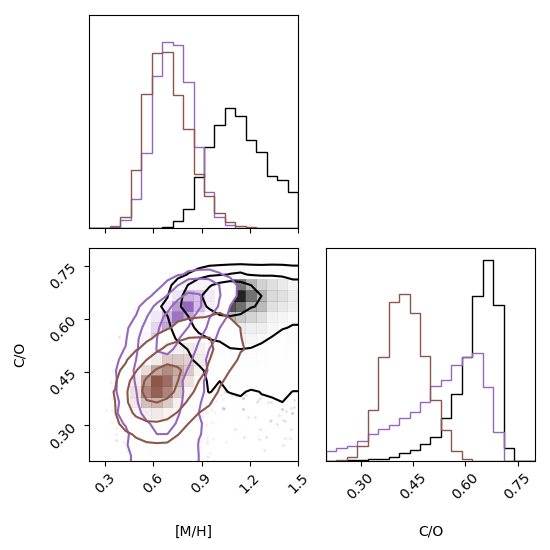}}\qquad\subfigure {\includegraphics
    [width=0.48\textwidth]{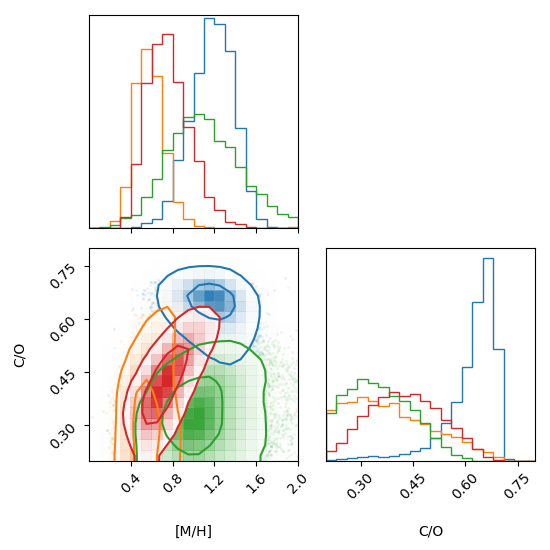}}
    \caption{Posterior distributions inferred from full equilibrium chemistry retrievals.  Left: Metallicity and C/O ratio inferred from the NIRCam transmission spectrum (purple), the NIRCam emission spectrum (brown), and the combined pre-JWST transmission and emission spectra (black).  Right: posterior distributions inferred from the F322W2 transmission spectrum (blue), F444W transmission spectrum (orange), F322W2 emission spectrum (green), and F444 emission spectrum (red).  In both subplots, 1$\sigma$ and 2$\sigma$ contours are plotted.  Note that the F444W transit posterior has a second mode at very high metallicity ($\sim$200x), which is outside the range of this plot.}
\label{fig:Z_CO_corners}
\end{figure*}

Figure \ref{fig:Z_CO_corners} (left) shows the metallicity and C/O ratios inferred from the transmission and emission spectra, compared to the posterior distribution inferred from pre-JWST (HST + Spitzer) data in \cite{zhang_2020}.  Numerical values for metallicity and C/O are in Table \ref{table:metallicity_CO_results}.  Both JWST-inferred posteriors are 2$\sigma$ consistent with the pre-JWST posteriors, and all three are 1$\sigma$ consistent with a metallicity of several times solar.  The C/O ratios are also consistent with each other to 2$\sigma$.  While the pre-JWST data prefers solar or super-solar C/O ratios, the NIRCam transit spectrum puts only loose constraints on C/O at the 3$\sigma$ level, and the emission spectrum has a slight preference for sub-solar C/O.
 
\begin{table*}[htbp]
    \centering
    \setlength{\tabcolsep}{6pt}
    \begin{tabular}{c C C C}
        \hline
        Data & \text{[M/H]} & \text{C/O} & \text{ln(z)}\\
	\hline
        HST+Spitzer transit+eclipse & 1.14_{-0.17}^{+0.25} & 0.64_{-0.07}^{+0.04} & 693.4\\
        F322W2+F444W transit & 0.72_{-0.11}^{+0.12} & 0.55_{-0.15}^{+0.09} & 463.2\\
        \textbf{F322W2+F444W eclipse} & 0.68_{-0.11}^{+0.15} & 0.43_{-0.05}^{+0.06} & 460.5\\
        F322W2 transit & 1.19_{-0.22}^{+0.19} & 0.64_{-0.08}^{+0.04} & 269.1 \\
        F444W transit & 0.67_{-0.19}^{+1.69} & 0.37_{-0.11}^{+0.15} & 191.5\\
        \textbf{F322W2 eclipse} & 1.10_{-0.33}^{+0.42} & 0.35_{-0.09}^{+0.11} & 273.7\\
        \textbf{F444W eclipse} & 0.74_{-0.18}^{+0.23} & 0.43 \pm 0.11 & 182.7\\
        \hline
        \textbf{HST+Spitzer transit+eclipse} & 1.14_{-0.17}^{+0.23} & 0.64_{-0.07}^{+0.04} & 693.3\\
        \textbf{F322W2+F444W transit} & 0.53_{-0.12}^{+0.13} & 0.41_{-0.12}^{+0.13} & 465.2\\
        F322W2+F444W eclipse & 0.59_{-0.10}^{+0.12} & 0.39 \pm 0.05 & 460.8\\
        \textbf{F322W2 transit} & 0.61_{-0.56}^{+0.37} & 0.60_{-0.11}^{+0.07} & 271.0\\
        \textbf{F444W transit} & 0.71_{-0.21}^{+1.66} & 0.39_{-0.12}^{+0.15} & 191.4\\
        F322W2 eclipse & 0.74_{-0.49}^{+0.48} & 0.41 \pm 0.12 & 274.8\\
        F444W eclipse & 0.73_{-0.16}^{+0.21} & 0.41_{-0.10}^{+0.11} & 182.7\\
        \hline
    \end{tabular}
    \caption{Metallicities and C/O ratios inferred from our equilibrium retrievals, along with the log Bayesian evidence of each retrieval.  The top half report retrievals that neglect methane depletion, while the retrievals in the bottom half account for methane depletion by fitting for an abundance multiplier.  In \textbf{bold} are the retrievals we adopt, namely the methane-depleted runs if transit data are included and the full equilibrium runs if only eclipse data are included.  We consider methane depletion because zeroing the methane opacity improves the fit to our transmission spectrum (Table \ref{table:molecules_test}), and because methane depletion is commonly observed in other exoplanet atmospheres (see Subsection \ref{subsec:methane_depletion}).} 
    \label{table:metallicity_CO_results}
\end{table*}

To gain insight into the constraining power of the two NIRCam bands individually, we perform separate retrievals for each band in each geometry.  The posteriors are plotted in Figure \ref{fig:Z_CO_corners} (right), and numerical values are given in Table \ref{table:metallicity_CO_results}.  The F444W transit and eclipse both prefer slightly higher (0.2--0.6 dex) metallicities than their F322W2 counterparts, but the wide posteriors mean that all values are consistent to 2$\sigma$.  All C/O ratio posteriors are very broad except the one produced by the F322W2 transit, which prefers higher ratios.  Despite this preference, all C/O posteriors are consistent with each other.  The F444W transit posterior is bimodal, with a second mode at 200x solar metallicity (hence the highly asymmetric [M/H] error bars in Table \ref{table:metallicity_CO_results}).

It is not clear why the F322W2 transit prefers higher C/O ratios than the other datasets.  This visit is likely the one most susceptible to stellar contamination--the eclipses are minimally affected by starspots and faculae, while the other transit is at a redder wavelength, where inhomgeneity-induced contrasts are lower.  In addition, a large starspot crossing was seen in the F322W2 transit data and the impacted data was removed (following \citealt{fu_2024}).  The excised data compromised 22\% of the transit, raising concerns about whether unocculted starspots or less obvious occulted starspots could still be contaminating the transit spectrum.  The F322W2 transit is also more susceptible to clouds and hazes than the other visits--the eclipses probe the hotter and probably less cloudy dayside, while the F444W transit is less impacted because Mie extinction cross sections decrease with wavelength for particles smaller than $\sim\lambda/(2\pi)$.  It is possible that a combination of stellar inhomogeneity and imperfectly modelled clouds are responsible for the high C/O ratio.


The metallicity values in Table \ref{table:metallicity_CO_results} (top half) span 4.7--15x solar, while the C/O ratios span 0.35--0.64.  Naively taking the mean values of the F322W2 + F444W NIRCam transit retrieval and the F322W2 + F444W NIRCam eclipse retrieval, we obtain a metallicity of 5x solar ([M/H]=0.70) and C/O=0.49.  Both the metallicity and the C/O are slightly ($\lesssim 2\sigma$) lower than in the pre-JWST retrieval.

\subsection{Metallicity and C/O: methane depletion}
\label{subsec:methane_depletion}

\begin{figure*}[ht]
  \centering \subfigure {\includegraphics
    [width=0.48\textwidth]{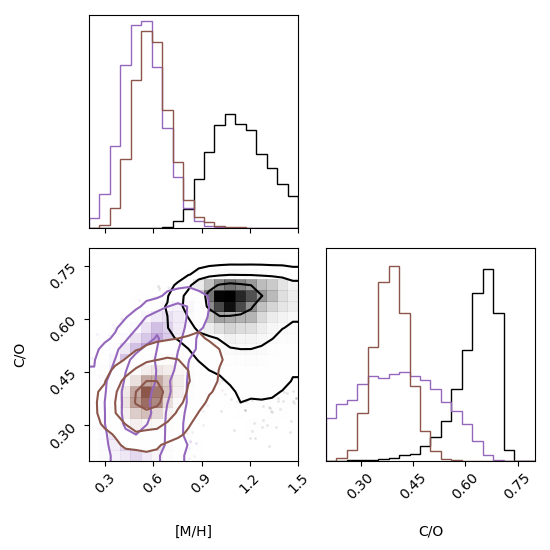}}\qquad\subfigure {\includegraphics
    [width=0.48\textwidth]{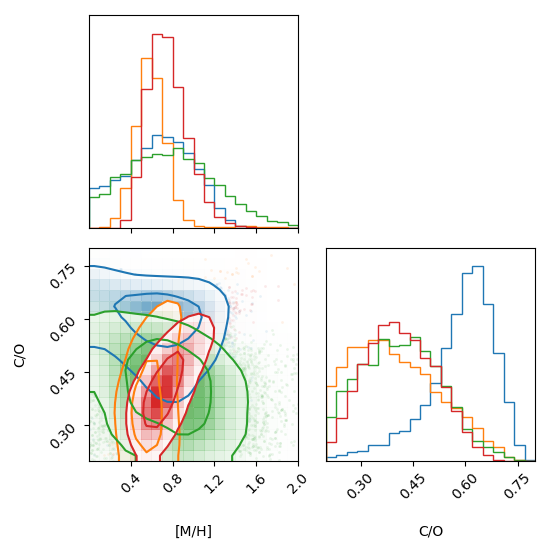}}
    \caption{Same as Figure \ref{fig:Z_CO_corners}, except that a methane multiple is now a free parameter in all retrievals in order to account for possible methane depletion.  For the joint pre-JWST transit and eclipse retrieval (left, black), we only applied the methane multiple to the transmission spectrum.  Just as for Figure \ref{fig:Z_CO_corners}, the F444W transit posterior has a second mode at very high metallicity ($\sim$200x), which is outside the range of this plot.}
\label{fig:Z_CO_corners_depCH4}
\end{figure*}

The equilibrium chemistry retrievals on the transmission spectrum revealed that methane depletion is preferred at a Bayes ratio of exp(465.2-463.2)=7.  Methane depletion would not be surprising, not only because \cite{fu_2024} inferred methane depletion using the same data, but because methane depletion has been observed time and again in exoplanet atmospheres \citep{stevenson_2010,kreidberg_2018,fu_2022,bell_2024,dyrek_2024}.  To determine how methane depletion would affect our metallicity and C/O ratio inferences, we reran all retrievals plotted in Figure \ref{fig:Z_CO_corners} with a methane multiple parameter $r$.  This parameter multiplies the equilibrium methane abundances.  We impose a uniform prior on log(r), with boundaries of -6 and +3.

Figure \ref{fig:Z_CO_corners_depCH4} shows how our posteriors change when methane depletion is taken into account.  On the left, the NIRCam transit and eclipse posteriors both shift to slightly lower metallicity and C/O ratio, with the former moving more than the latter.  The pre-JWST posterior broadens but does not shift significantly, making it slightly less consistent with the NIRCam posteriors.  On the right, the most dramatic change occurs for the F322W2 posteriors, which dramatically broaden, especially in emission.  The F322W2 transit posterior moves to slightly lower metallicity and C/O ratio; this, combined with its larger width, completely eliminates the tension with the other three retrievals.

Accounting for methane depletion decreases the inferred log metallicity by 0.19/0.09 and the C/O ratio by 0.14/0.04 (transmission/emission), as shown in Table \ref{table:metallicity_CO_results}.  In transmission, the inferred methane multiple is consistent with the lower end of the prior ($10^{-6}$, i.e. depletion by a factor of a million), and its 2$\sigma$ upper limit is $10^{-1.6}$.  However, the multiple is also consistent with no depletion at 3$\sigma$.  In emission, the inferred methane multiple is 2$\sigma$ consistent with both the lower end of the prior ($10^{-6}$) and with 1.  The Bayesian evidence favors the inclusion of the methane depletion factor for transmission ($\Delta$ln(z)=2.0), but does not strongly favor either for emission ($\Delta$ln(z)=0.3).  We therefore adopt as fiducial the transmission retrievals with methane depletion and the emission retrievals without.

\begin{table}[htbp]
    \centering
    \setlength{\tabcolsep}{6pt}
    \begin{tabular}{c C C C C}
        \hline
        Cloud & \text{[M/H]} & \text{C/O} & \text{ln(z)} & \chi^2 \\
	\hline
        Mie & 0.42_{-0.10}^{+0.11} & 0.35_{-0.10}^{+0.15} & 463.8 & 58.0\\
        Fixed Mie & 0.44_{-0.11}^{+0.12} & 0.38_{-0.11}^{+0.16} & 463.7 & 63.1\\
        Clear & 0.42_{-0.11}^{+0.12} & 0.36_{-0.11}^{+0.15} & 463.7 & 63.5\\
        \textbf{Power law} & 0.53_{-0.12}^{+0.13} & 0.41_{-0.12}^{+0.13} & 465.2 & 47.5\\
        Opaque & 0.46_{-0.11}^{+0.12} & 0.36_{-0.10}^{+0.14} & 463.8 & 57.8\\
        \hline
    \end{tabular}
    \caption{Comparison of different cloud parameterizations for the transmission spectrum retrieval including methane depletion.  ln(z) is the natural log of the Bayesian evidence.  The ``power law'' parameterization also includes an opaque cloud deck.}
    \label{table:alternate_cloud_params}
\end{table}

With methane depletion enabled, we explored the effect of differing aerosol parameterizations on the transit spectrum retrieval.  Aside from the fiducial approach of a power law extinction cross section plus an opaque cloud deck, we adopted a clear atmosphere, an opaque cloud deck on its own, Mie scattering with spherical particles from a log-normal radius distribution (as adopted in Z20), and Mie scattering with parameters fixed to the best fit values from Z20.  All approaches resulted in similar values for [M/H] and C/O, as shown in Table \ref{table:alternate_cloud_params}.  The fiducial approach results in the lowest $\chi^2$ and highest Bayesian evidence, which is why we adopted it as the fiducial approach.

\subsection{Molecular detections}
An equilibrium retrieval does not automatically answer the question of which molecules are present in the atmosphere.  To answer this question, we use one of the features introduced in the new version of PLATON: an opacity zeroer, which makes a user-specified list of molecules completely transparent to light.  First, we take the best fit transit/eclipse spectrum from the full equilibrium retrieval and remove opacity from different molecules to probe their effect on the spectrum, keeping all other parameters the same.  If the removal of a certain molecule appreciably changes the model spectrum such that it is no longer consistent with the data, we run a retrieval with that molecule's opacity zeroed to see if an alternate set of atmospheric parameters can explain the observations equally well.  If so, the molecule cannot be considered detected; if not, and if the Bayesian evidence strongly disfavors the retrieval with the zeroed opacity, we can claim a molecular detection.

\begin{figure*}[ht]
  \centering \subfigure {\includegraphics
    [width=0.48\textwidth]{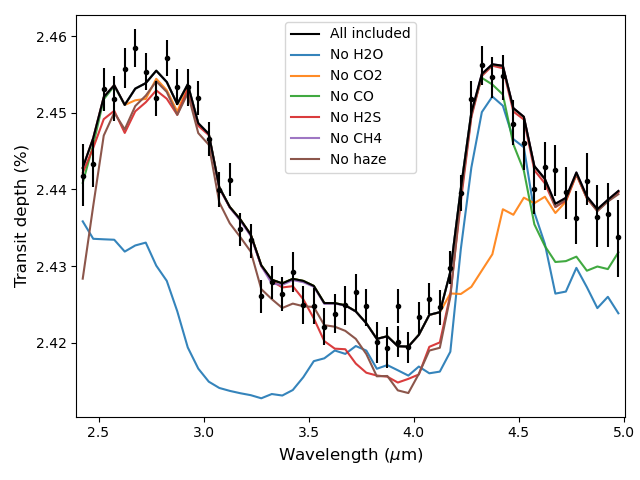}}\qquad\subfigure {\includegraphics
    [width=0.48\textwidth]{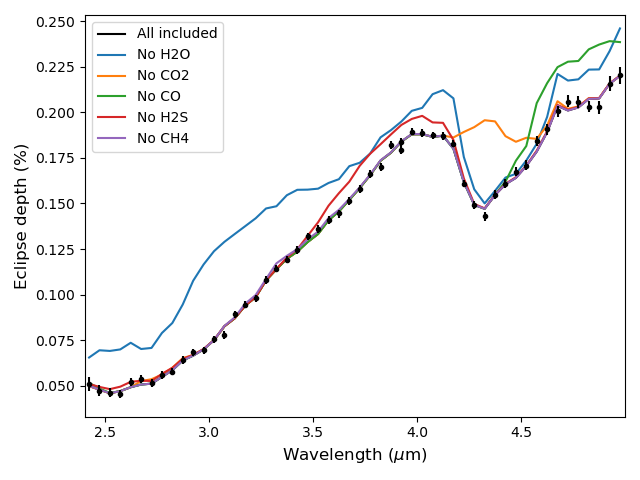}}
    \caption{Best fit transmission (left) and emission (right) spectra of HD 189733b, from four NIRCam transit observations.  The black line indicates the best fit model from PLATON retrievals, while the colored lines indicate the effect of removing a single opacity source (e.g. the opacity of a single molecule) from the best fit model.}
\label{fig:best_fits}
\end{figure*}

Even though we include 11 species in our retrievals (C$_2$H$_2$, CH$_4$, CO$_2$, CO, H$_2$O, H$_2$S, HCN, K, Na, NH$_3$, SO$_2$), not all contribute significant opacity in our best fit model.  In the transmission spectrum, only H$_2$O, CO$_2$, CO, H$_2$S, and CH$_4$ contribute significant opacity: zeroing all other molecular opacities changes $\chi^2$ by less than 0.01 from its fiducial value of 50.9.  In order of decreasing impact on $\chi^2$, the most important opacity sources are H$_2$O, CO$_2$, the haze, H$_2$S, CO, and CH$_4$. It is intriguing that removing CH$_4$ improves the quality of the fit by $\Delta \chi^2=3.3$, while removing any of other absorbers dramatically worsens the fit.  Figure \ref{fig:best_fits} shows the impact on the best fit transit spectrum of removing each absorber.  H$_2$O and CO$_2$ show extremely strong and obvious absorption peaks, so their existence in the atmosphere is not in doubt.  CO raises the transit spectrum redward of 4.4 $\mu$m.  H$_2$S turns what would otherwise be a dip in the transit spectrum from 3.5--4.1 $\mu$m into a plateau.  In comparison, CH$_4$ only has a very subtle effect on the spectrum, slightly raising the transit depths around 3.4 $\mu$m.

In our transmission retrievals where the opacity of one species was zeroed, we find that the Bayesian evidence significantly decreased ($\Delta \log{z} < -6.1$) unless that species was CH$_4$, in which case it increased by $\Delta \log{z} = 2.8$ (see Table \ref{table:molecules_test}).  We thus confirm the results of \cite{fu_2024}: our transit retrievals strongly favor the existence of H$_2$O, CO$_2$, CO, and H$_2$S, while indicating that methane is depleted from equilibrium abundances.

The emission spectrum corroborates the detection of H$_2$O, CO$_2$, CO, and H$_2$S in the transmission spectrum.  It is consistent with both an equilibrium-level methane abundance, and with no methane at all.

\begin{table}[htbp]
    \centering
    \setlength{\tabcolsep}{6pt}
    \begin{tabular}{c c C}
        \hline
        Geometry & Model & \ln{z} \\
	\hline
        Transit & Fiducial & \textbf{463.2}\\
        Transit & No haze  & 456.8 \\
        Transit & No H$_2$O   & 428.5 \\
        Transit & No CO$_2$   & 423.2\\
        Transit & No CO    & 457.1\\
        Transit & No H$_2$S   & 456.0\\        
        Transit & No CH$_4$   & 466.0\\
        \hline
        Eclipse & Fiducial & \textbf{460.5}\\
        Eclipse & No H$_2$O   & 388.5\\
        Eclipse & No CO$_2$   & 397.3\\
        Eclipse & No CO    & 428.2\\        
        Eclipse & No H$_2$S   & 458.7\\
        Eclipse & No CH$_4$   & 460.5\\
        \hline
    \end{tabular}
    \caption{The impact of zeroing the opacity of different molecules on the fit quality, setting all parameters to the best fit values. To see how strongly a certain molecule is detected, compare the corresponding log Bayesian evidence ln(z) column with the ln(z) of the fiducial retrieval.}
    \label{table:molecules_test}
\end{table}

\subsection{The effect of photochemistry}
One possible physical cause of methane depletion is quenching: movement of air from hotter and less methane-rich regions to the terminator's transmission photosphere.  Another possible cause is photochemistry, which begins with the breakdown of molecules such as CH$_4$ and H$_2$S into their constituent atoms.  These atoms proceed to form hydrocarbon hazes, SO2, and other photochemical products.

To explore the effect of photochemistry on our HD 189733b retrievals, we run the open-source 1D photochemical kinetics code VULCAN \citep{tsai_2021} to obtain molecular abundance profiles with and without photochemistry.  VULCAN comes with an example configuration file for HD 189733b, which was used by \cite{tsai_2021} in their study.  The temperature-pressure profile and Kzz profile that the example file adopts are from GCMs run by \cite{moses_2011}.  We make four modifications to the file: we add sulfur to the list of atoms, which by default includes only H, O, C, and N; we use the SNCHO photochemistry network instead of the NCHO photochemistry network; we adopt the high-energy stellar spectrum from \cite{bourrier_2020} instead of \cite{moses_2011} (both spectra are included with VULCAN); we change the metallicity from solar to 6x solar; and we change the solar abundances from the default \cite{lodders_2009} to the newer and slightly different \cite{asplund_2021}.

\begin{figure}[ht]
  \centering \includegraphics
    [width=0.5\textwidth]{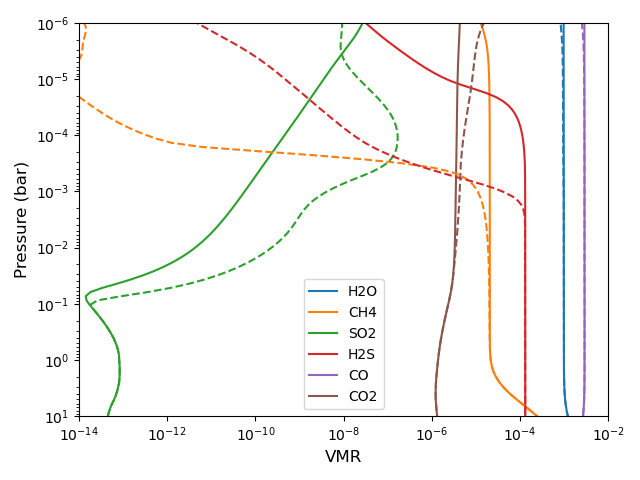}
    \caption{Abundances of selected molecules as predicted by VULCAN, with photochemistry (dashed lines) and without (solid lines).}
\label{fig:photochem_abunds}
\end{figure}

Figure \ref{fig:photochem_abunds} shows the effects of photochemistry on the abundances of selected species.  H$_2$O and CO have such high abundance that photochemistry has a negligible effect.  CO$_2$ becomes slightly more abundant at low pressures due to photochemistry.  SO2 becomes significantly more abundant, rising from 1 ppb to 1 ppm at 0.1 mbar.  CH$_4$ and H$_2$S virtually disappear from the atmosphere at pressures lower than 1 mbar, but are minimally affected deeper in the atmosphere.

To crudely mimic the effects of photochemistry in our retrieval, we zero out the abundance of both H$_2$S and CH$_4$ at pressures below 1 mbar while assuming equilibrium abundances at all other pressures.  After doing so, the log metallicity of the transmission spectrum retrieval falls to $0.51 \pm 0.12$, and the C/O ratio falls to $0.40_{-0.12}^{+0.13}$ (compare to the ``F322W2+F444W transit'' results in the top half of Table \ref{table:metallicity_CO_results}). The C/O posterior is wide, with a 2$\sigma$ range of 0.30 to 0.75.  The log metallicity of the emission retrievals falls to $0.76_{-0.14}^{+0.15}$ and the C/O ratio to $0.34_{-0.06}^{+0.08}$ (compare to the ``F322W2+F444W'' values in the top half of Table \ref{table:metallicity_CO_results}).  The transmission retrieval results are strikingly similar to what we obtained in Subsection \ref{subsec:methane_depletion} by adding a free parameter to model methane depletion.  They suggest that for HD 189733b, photochemistry is an important cause of methane depletion.

The effect of photochemistry depends on the temperature-pressure profile, chemical composition, Kzz, and XUV spectrum, among other factors.  Kzz and the XUV spectrum have especially large uncertainties.  This subsection should not be considered a definitive study of HD 189733b's photochemistry, but a qualitative order-of-magnitude spot test of the importance of photochemistry and its impact on retrievals.

\subsection{The temperature-pressure profile and its degeneracy with C/O ratio}
\begin{figure*}[ht]
  \centering 
  \subfigure {\includegraphics
    [width=0.48\textwidth]{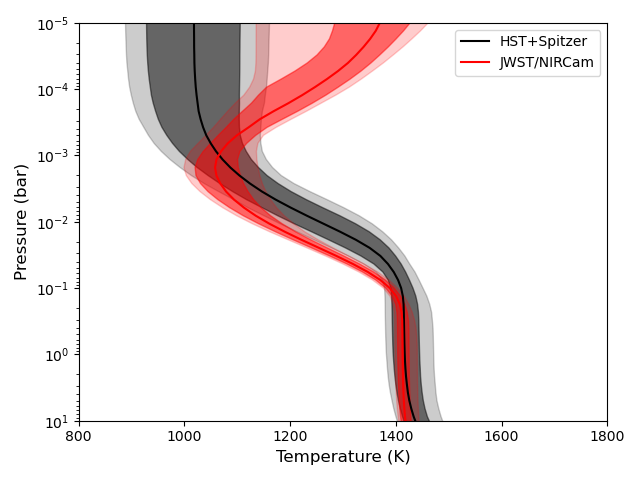}}\qquad\subfigure {\includegraphics
    [width=0.48\textwidth]{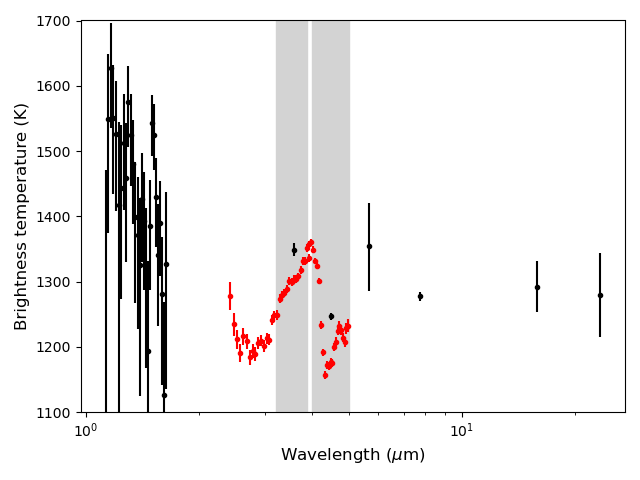}}
    \caption{Left: temperature-pressure profile of HD 189733b inferred from NIRCam emission spectra (red), compared with that inferred from HST and Spitzer emission spectra (black).  Right: the emission spectrum used to infer the T/P profile, converted to brightness temperature.  The gray bands indicate the bandpasses of Spitzer/IRAC channels 1 and 2, which overlap with the NIRCam data.  At a given wavelength, the brightness temperature is roughly the temperature of the photosphere.  It can be compared to 1540 K, the temperature of a zero albedo, zero heat redistribution blackbody at HD 189733b's semimajor axis.}
\label{fig:TP_profile}
\end{figure*}

The reason that emission spectra show any features at all is because the dayside atmosphere is non-isothermal.  The emission spectrum is therefore sensitive to the temperature-pressure profile of the atmosphere.  Figure \ref{fig:TP_profile} shows the inferred T/P profile from the NIRCam emission spectrum, compared to that inferred from the pre-JWST (HST+Spitzer) retrieval in \cite{zhang_2020}.  The profiles are consistent to 2-3$\sigma$ at all pressures, but the NIRCam data has a 1.8$\sigma$ preference for a thermal inversion around 2 mbar, reflected by the posterior distribution of log($\gamma_2$) favoring values above 0.  Whether a thermal inversion is physically plausible at 2 mbar is an open question.  One might argue that thermal inversions are implausible because they have not been seen on HD 189733b despite extensive observational data at both low and high resolution; because thermal inversions are not theoretically expected for planets with $T_{eq}$ below 1500-2000 K (c.f. \citealt{gandhi_2019} and its references); and because thermal inversions have proven observationally elusive for planets below $T_{eq}=$2000 K \citep{baxter_2020,mansfield_2021,changeat_2022}.  However, all of these arguments have weaknesses.  The lack of a thermal inversion spanning the photosphere is immediately obvious in emission spectra--such an inversion would turn absorption features into emission features--but as our retrieval shows, this does not mean that there is no thermal inversion higher in the atmosphere.  Theoretical understanding of thermal inversions is still poor, and although the theory has tended to over-predict the prevalence of inversions, there are plausible mechanisms for generating inversions in planets colder than ultra-hot Jupiters at very low pressures (e.g. atomic sodium, as proposed by \citealt{sing_2008}).

In our emission retrievals, we see a degeneracy between $\gamma_2$ and C/O: inverted atmospheres ($\log(\gamma_2) > 0$) prefer lower C/O ratios than non-inverted atmospheres.  We also see a degeneracy between $\gamma_2$ and [M/H]: inverted atmospheres prefer lower metallicities.  We experimented with running an emission spectroscopy retrieval that disallows thermal inversions by forcing $\log(\gamma) < 1$ and $\log(\gamma_2) < 1$ and obtained [M/H]=$0.98 \pm 0.10$ and C/O=$0.53 \pm 0.04$, which are higher than the fiducial retrieval's values by 0.30 and 0.10 respectively.

\subsection{Free retrievals}
In addition to the equilibrium retrievals, we performed free retrievals on the transmission and emission spectra.  These retrievals have the same setup as their equilibrium chemistry counterparts, except that [M/H] and C/O are switched out for the log VMRs of CH$_4$, CO$_2$, CO, H$_2$O, and H$_2$S: the five molecules with a non-negligible effect on the spectrum in our equilibrium best fit solution.  The log VMR parameters were given flat priors from -12 to -0.5, and the mixing ratio left over after summing the VMRs is attributed to a mixture of H2 and He.

For each sample in the equally weighted posterior, we compute a metallicity from the free retrievals by summing up the number of metal atoms (C, O, S), dividing it by the number of H atoms (including those found in CH$_4$, H$_2$O, and H$_2$S), and comparing it to the solar value of (C+O+S)/H=8.6\e{-4}.  The C/O ratio is found simply by dividing the number of C atoms by the number of O atoms.  These numbers are still not directly comparable to the equilibrium retrievals because they are gas-phase metallicity and C/O, whereas the equilibrium metallicity and C/O include condensates; and less importantly, because they do not include metals in other molecules.  To address these problems, we sample from the equilibrium retrieval chains, compute the atmospheric mixing ratios at the approximate location of the photosphere (1 mbar in transit, 20 mbar in emission), and convert these mixing ratios to an adjusted metallicity and C/O with the exact same approach as we use for the free retrievals.

\begin{figure*}[ht]
  \centering 
  \subfigure {\includegraphics
    [width=0.48\textwidth]{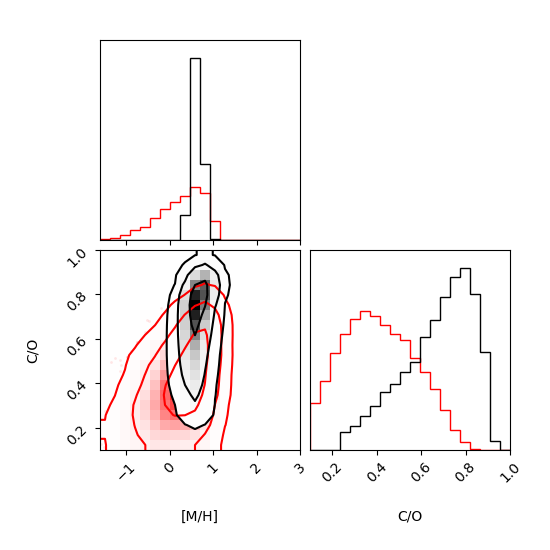}}\qquad\subfigure {\includegraphics
    [width=0.48\textwidth]{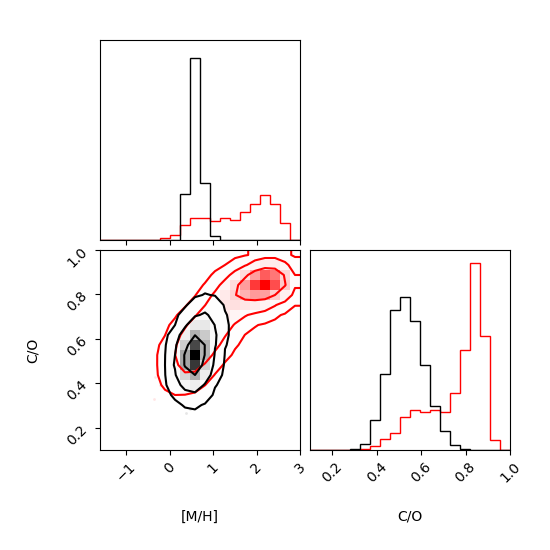}}
    \caption{Comparison between the adjusted gas-phase metallicity and C/O inferred from the equilibrium retrieval (black) and the free retrieval (red), for transit (left) and eclipse (right).}
\label{fig:free_equi_comp}
\end{figure*}

Figure \ref{fig:free_equi_comp} compares the adjusted gas-phase metallicity and C/O inferred from the free retrievals to those inferred from the equilibrium retrievals.  Consistent with the findings of \cite{fu_2024}, the transmission free retrieval is fully consistent with the equilibrium retrieval, but with extremely broad posteriors in both parameters.  The emission retrieval prefers extremely high metallicites, but the posterior has a very long 2$\sigma$ tail that encompasses the peak of the equilibrium posterior.  We conclude that for NIRCam transit and eclipse spectra, free retrievals have too many degrees of freedom to give robust results.  These degrees of freedom result in extremely broad posteriors even if the results can be trusted, limiting the usefulness of free retrievals for these datasets.

\section{Discussion}
\label{sec:discussion}
HD 189733b is one of the most extensively studied exoplanets, and its chemical composition has been constrained at both low and high resolution, in both transmission and emission.  In the previous section, we compared our JWST retrieval results to our pre-JWST retrieval results.  Here, we compare to the results obtained by other teams, summarized in Table \ref{table:retrieval_results}.

\cite{fu_2024} used the same NIRCam transit data as we did, but their fiducial analysis used a different reduction package (written by G. Fu) and their retrievals were not performed with PLATON.  First, they performed a 1D radiative-convective-photochemical equilibrium grid retrieval by coupling the Sc-CHIMERA RCE solver with the VULCAN photochemical kinetics solver \citep{tsai_2017}.  As is common with grid retrievals, they obtained very narrow posteriors for both metallicity ([M/H] = $0.57_{-0.04}^{+0.05}$) and C/O (C/O $<$ 0.2 at 5$\sigma$).  Our result is consistent with theirs when we account for either methane depletion or photochemistry, because our C/O posteriors have a long tail toward low C/O.  In addition to the 1D radiative-convective-photochemical-equilibrium (RCPE) grid retrieval, \cite{fu_2024} also performed a free retrieval using CHIMERA \citep{line_2013}.  Unlike us, they adopt a non-isothermal T/P profile and a parameterization of inhomogeneous clouds.  The mixing ratio posteriors imply C/O=$0.27_{-0.09}^{+0.10}$ and a poorly constrained metallicity of $-0.37_{-0.47}^{+0.80}$, consistent with their grid retrieval.  We obtain similarly poorly constrained posteriors with our transmission free retrieval.

\begin{table*}[htbp]
    \centering
    \setlength{\tabcolsep}{6pt}
    \begin{tabular}{c c c c C C C C c}
        \hline
        Geometry & Instrument & $\lambda$ (\um) & Type & \text{[M/H]} & \text{C/O} & \text{log(H2O)} & \text{log(CO)} & Ref\\
	\hline
        Transit & NIRCam & 2.7--5.3 & Equilib & 0.53_{-0.13}^{+0.12} & 0.41_{-0.13}^{+0.12} & -2.97_{-0.21}^{+0.19} & -2.98_{-0.19}^{+0.22} & Z24\\
        Transit & NIRCam & 2.7--5.3 & 1D RCPE & 0.57_{-0.04}^{+0.05} & <0.2 (5\sigma) & &  & F24\\
        Transit & NIRCam & 2.7--5.3 & Free & -0.37_{-0.47}^{+0.80} & 0.27_{-0.09}^{+0.10} & -3.69_{-0.46}^{+0.76} & -4.13_{-0.56}^{+0.88} & F24\\        
        Transit & CARMENES & 0.96--1.71 & Free & 1? & 0.5? & -2.39_{-0.12}^{+0.16} & < -3.71 & B24\\

        Transit & SPIRou & 0.9--2.5$^a$ & Free & & & {-3.84_{-0.53}^{+0.75}}^b & & K24\\
        Transit & SPIRou & 0.9--2.5$^a$ & Free & & & -4.4 \pm 0.4 & & B21\\
        Transit & Many$^c$ & 0.3--5.0$^a$ & Free & & & -5.04_{-0.30}^{+0.46} & -6.9_{-3.3}^{+3.0} & P19 \\       
        Transit & Many$^d$ & 0.3--24 & Free & & & >-5.7, <-2.7 & & L14\\
        \hline
        Eclipse & NIRCam & 2.7--5.3 & Equilib & 0.68_{-0.11}^{+0.15} & 0.43_{-0.05}^{+0.06} & -2.67 \pm 0.14 & -2.50_{-0.27}^{+0.20} & Z24\\        
        Eclipse & KPIC & 2.2--2.5$^a$ & Free & \text{[C/H]}=0.4 \pm 0.5 & 0.3 \pm 0.1 & -2.9 \pm 0.4$^b$ & -3.3 \pm 0.5$^b$ & Fi24\\
        & & & & \text{[O/H]} = 0.8 \pm 0.4 & & & & \\
        Eclipse & Many$^e$ & 1.45--24 & Free &  & 0.45-1.0 & >-4.0,<-2.3 & & L12\\
        \hline
        Combined & Many$^f$ & 0.3--26 & Equilib & 1.08_{-0.23}^{+0.22} & 0.66_{-0.09}^{+0.05} & &  & Z20\\
        Combined & Many$^f$ & 0.3--26 & Equilib & 1.07_{-0.18}^{+0.23} & 0.64_{-0.08}^{+0.05} & &  & Z24\\
    \end{tabular}
    \caption{Published retrievals on HD 189733b transmission or emission spectra from 2012 or later, compared to the fiducial results of this work, which include methane depletion for transit but not eclipse. $^?$B24 find that their water mixing ratio is consistent with 10x solar metallicity and C/O=0.5, but also with 3x solar metallicity and C/O=0.1, and with a wide range of other combinations} $^a$Not continuous.  $^b$Converted from mass mixing ratio by assuming $\mu=2.3$. $^c$ STIS, ACIS, WFC3, IRAC. $^d$ STIS,ACS,NICMOS,WFC3,IRAC,MIPS $^e$ NICMOS,IRS,IRAC,MIPS $^f$ STIS,WFC3,IRAC,MIPS. References: Z24=this work; F24=\cite{fu_2024}; B24=\cite{blain_2024}; K24=\cite{klein_2024}; B21=\cite{boucher_2021}; P19=\cite{pinhas_2019}; L14=\cite{lee_2014}; Fi24=\cite{finnerty_2024}; L12=\cite{lee_2012}
    \label{table:retrieval_results}
\end{table*}

HD 189733b has been observed with high-resolution spectroscopy in both transit and eclipse.  \cite{blain_2024} used CARMENES transmission spectra to derive a water mixing ratio 0.6 dex higher than our transmission equilibrium retrieval's inferred abundance at 1 mbar (roughly the location of the photosphere in transit), consistent with 10x solar metallicity and solar C/O.  However, their non-detection of CH$_4$, CO, H$_2$S, HCN, and NH3 is consistent only with sub-solar metallicity.  Alternatively, it is consistent with a very low C/O ratio of 0.1.  \cite{klein_2024} used SPIRou transmission spectra to measure a water abundance 0.9 dex lower than, but consistent with, our result.  \cite{klein_2024} was a re-analysis of the data first published by \cite{boucher_2021}, who obtained a substantially lower abundance of log(H$_2$O)=$-4.4 \pm 0.4$.  \cite{klein_2024} used a new open-source pipeline that they developed, but it is not clear which of the differences with \cite{boucher_2021} are responsible for the different abundance.

Emission spectroscopy retrievals of HD 189733b are not as numerous.  The most recent low-resolution retrievals are the one in this work and our pre-JWST retrieval on HST and Spitzer data.  \cite{finnerty_2024} observed high-resolution emission spectra of the planet in K band using the Keck Planet Imager and Characterizer (KPIC) and obtained a log$_{10}$ volume mixing ratio of $-2.9 \pm 0.4$ for water and $-3.3 \pm 0.5$ for CO, with all other species undetected.  The water mixing ratio is fully consistent with our equilibrium result, and the CO mixing ratio is lower by 0.8 dex, which is consistent to 2$\sigma$.  They derive [C/H]=$0.4 \pm 0.5$ and [O/H]=$0.8 \pm 0.4$, both fully consistent with our retrieved metallicity.

A few older low-resolution retrievals exist in the literature.  These include \cite{pinhas_2019}, which combined STIS, ACS, WFC3, and IRAC transmission data to obtain a subsolar H$_2$O VMR of log(H$_2$O)=$-5.04_{-0.30}^{+0.46}$.  An earlier transmission retrieval, which additionally included NICMOS and MIPS data \citep{lee_2014}, found that depending on the assumed aerosol properties, log(H$_2$O) could range from -5.7 to -2.7.  In emission, \cite{lee_2012} performed a retrieval on spectra from HST/NICMOS, Spitzer/IRS, Spitzer/IRAC, and Spitzer/MIPS to obtain $-4.0 <$ log(H$_2$O) $< -2.3$, $-3.5 <$ log(CO$_2$) $< -1.8$, and log(CH4) $< -4.4$.  These together imply C/O=0.45--1.0.  Both ACS and NICMOS are no longer commonly used for exoplanet atmospheric research, having been made obsolete by WFC3.  In particular, the reliability of spectroscopic NICMOS observations of HD 189733b has been called into question \citep{gibson_2011,deming_seager_2017}.  

We summarize these transmission and emission retrieval results in Table \ref{table:retrieval_results}.  We chose not to include results earlier than 2012 due to the substantial advances in detector technology, data reduction techniques, line lists, and Bayesian inference methods that have occurred since then.  Examining the results we did include, we see that the water abundance is consistent with a metallicity of several times solar in the vast majority of the retrievals.  This is especially true for results published after 2021.  HD 189733b joins a number of other gas giants whose JWST spectra indicate a metallicity around 10x solar, including WASP-39b \citep{tsai_2023}, HD 209458b \citep{xue_2024}, and WASP-107b \citep{welbanks_2024}.

If the metallicity is relatively robust, the C/O ratio is less so.  Our own transmission spectrum retrievals favor a wide range of C/O ratios, ranging from sub-solar to super-solar, and adding a methane depletion factor makes C/O ratios less than $\sim$0.2 consistent with the transit data (although not the eclipse data).  Not accounting for methane depletion pushes up the C/O ratio, as does neglecting condensates and only counting C and O atoms in gas phase.  Literature C/O values include supersolar or near-solar ratios \citep{zhang_2020,lee_2012}, somewhat subsolar ratios ($\sim0.3$; \citealt{finnerty_2024,fu_2024}) and extremely subsolar ratios ($<0.1$; \citealt{fu_2024}).  The lack of robustness in C/O ratio constraints bodes poorly for the exoplanet field's long-standing goal of using this ratio to constrain planet formation and migration.

\subsection{Comparison with WASP-39b and HD 209458b}
\begin{figure}[h]
  \includegraphics
    [width=0.48\textwidth]{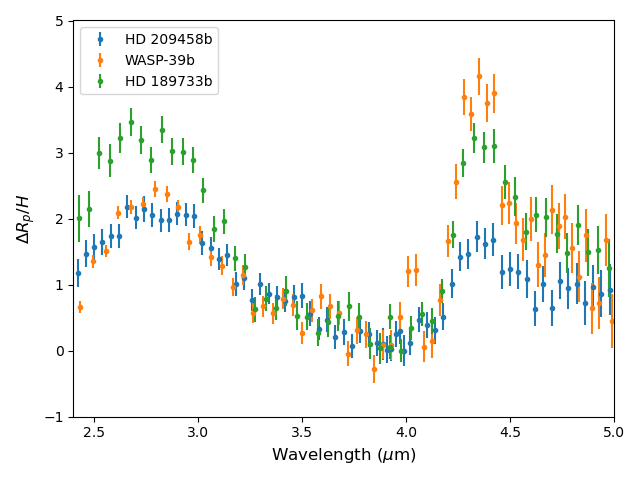}
    \caption{JWST transmission spectra of HD 189733b compared to that of two other hot Jupiters, all normalized by scale height.  HD 209458b and 189733b data are from NIRCam, while WASP-39b data is from PRISM.}
\label{fig:planets_comparison}
\end{figure}

Two other hot Jupiters of roughly similar equilibrium temperature have so far been observed in transmission with JWST in the same wavelength range: WASP-39b and HD 209458b.  Figure \ref{fig:planets_comparison} compares their transmission spectra, normalized by their respective scale heights.  We obtained the WASP-39b PRISM spectrum from \cite{rumstamkulov_2023} and the HD 209458b NIRCam spectrum from \cite{xue_2024}.

One striking difference between the planets is the relative height of their water and carbon dioxide peaks.  While WASP-39b's sharp CO$_2$ peak rises almost 2 scale heights above the water peak, HD 209458b and 189733b both have carbon dioxide peaks slightly below the water peak.  Since the relative height is most strongly affected by the relative abundance of the two species, and since high metallicity favors CO$_2$ more than it favors water, one might suspect that WASP-39b has a higher metallicity than the other two planets.  However, retrievals on all three planets obtain metallicities of $\sim$several times solar.  The retrieval by \cite{xue_2024} finds that the difference in relative peak height is instead explained by HD 209458b's low C/O ratio of $\sim$0.1.

To see why the HD 209458b data favors a low C/O ratio instead of a low metallicity, it is useful to calculate the equilibrium abundances of several species at 1200 K and 1 mbar, roughly the conditions at the photosphere.  In a [M/H]=0.5, C/O=0.1 atmosphere at this pressure and temperature, VMR(CO$_2$)/VMR(H$_2$O) = 3\e{-4}.  The same abundance ratio is found in a [M/H]=-0.05, C/O=0.5 atmosphere under the same conditions.  Sure enough, PLATON forward models of these two atmospheres show a very similar relative height between the two molecular peaks.  However, the models do not look similar redward of the CO$_2$ peak.  While the higher-metallicity C/O=0.1 atmosphere has VMR(CO)/VMR(CO$_2$) = 400, the lower-metallicity C/O=0.5 atmosphere has VMR(CO)/VMR(CO$_2$)=5000, more than an order of magnitude higher.  This raises the transmission spectrum redward of 4.5 \um (see Figure \ref{fig:best_fits} for intuition), which turns the CO$_2$ peak into more of a plateau.  Also, VMR(H$_2$S)/VMR(H$_2$O) = 0.025 for the former scenario, compared to 0.086 for the latter scenario.  The increased importance of H$_2$S fills in the opacity window around 3.5--4.0 \um.  Thus, it is the relative lack of CO and H$_2$S absorption in the HD 209458b spectrum that favors a low C/O ratio over a low metallicity.

HD 189733b also has a low CO$_2$ peak relative to its water peak.  Why does the retrieval not favor a low C/O ratio, as it did for HD 209458b?  The simplest explanation is that the retrieval believes that it sees H$_2$S and CO in the spectrum.  To get some insight into why this is, we use the Bayesian leave-one-out cross-validation method of estimating the expected log pointwise predictive density (elpd\textsubscript{LOO}), which was recently implemented in PLATON.  Specifically, we run a retrieval with the C/O fixed to 0.1, and calculate elpd\textsubscript{ref} - elpd\textsubscript{C/O=0.1} for every data point.  We then compare to elpd\textsubscript{ref} - elpd\textsubscript{no H2S} and elpd\textsubscript{ref} - elpd\textsubscript{no CO}.

\begin{figure*}[ht]
  \centering 
  \subfigure {\includegraphics[width=0.48\textwidth]{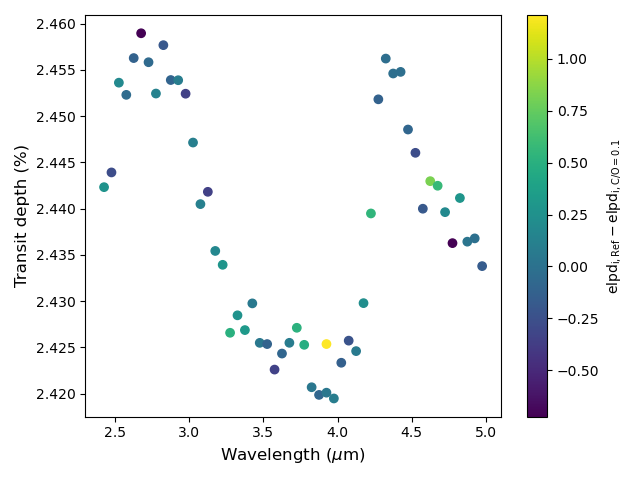}}\qquad\subfigure {\includegraphics[width=0.48\textwidth]{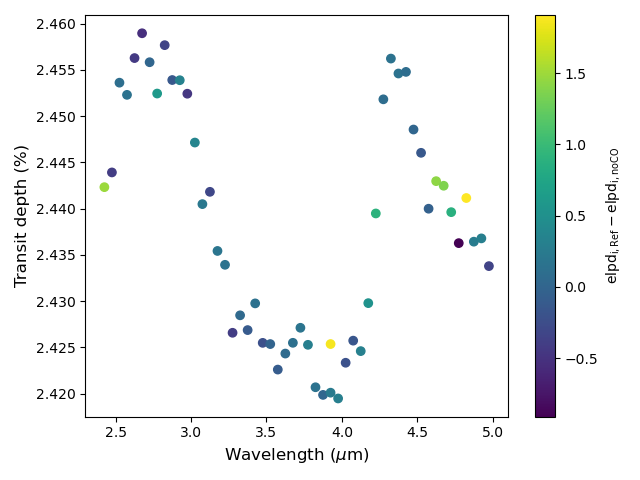}}
  \subfigure {\includegraphics[width=0.48\textwidth]{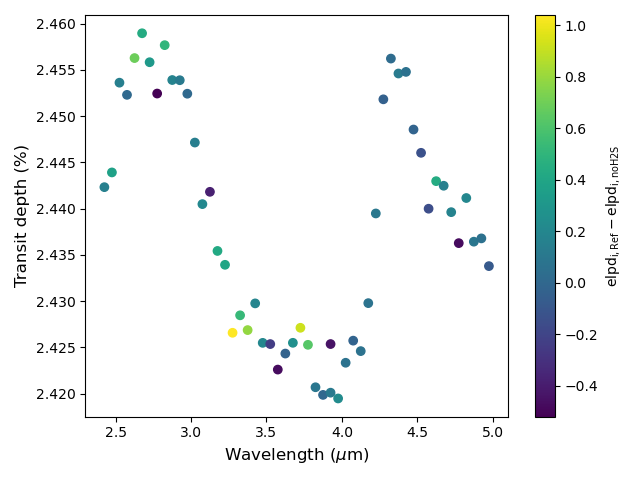}}\qquad\subfigure {\includegraphics[width=0.48\textwidth]{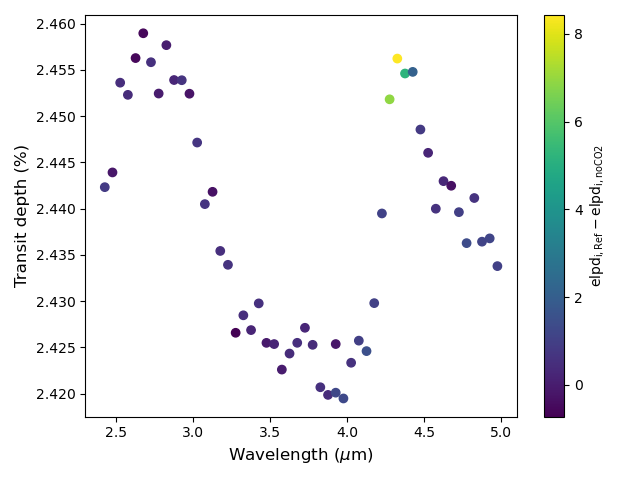}}
    \caption{Difference in expected log predicted density between the reference retrieval on HD 189733b (with a power law haze and no methane depletion) and an alternate retrieval.  The alternate retrievals have C/O fixed to 0.1 (upper left), or the opacity of one molecule zeroed out (upper right: CO; lower left: H$_2$S; lower right: CO$_2$).}
\label{fig:loo}
\end{figure*}

Figure \ref{fig:loo} shows these $\Delta$elpd values.  These is a strong resemblance between elpd\textsubscript{ref} - elpd\textsubscript{C/O=0.1} and elpd\textsubscript{ref} - elpd\textsubscript{no CO}, indicating that the detection of CO in part explains why the reference retrieval disfavors ultra-low C/O ratios.  This cannot be the whole story, however, because methane must play a role: our transmission retrieval with methane depletion is more consistent with very low C/O ratios (Figure \ref{fig:Z_CO_corners_depCH4}).  In addition, both the elpd\textsubscript{ref} - elpd\textsubscript{no CO} and elpd\textsubscript{ref} - elpd\textsubscript{no H2S} plots are not satisfying because they do not intuitively explain \textit{why} PLATON believes it sees CO/H$_2$S, only that the detection depends on many data points.  $\Delta$elpd plots are not always so inscrutable.  For example, the lower right panel of Figure \ref{fig:loo} shows elpd\textsubscript{ref} - elpd\textsubscript{no CO$_2$}, which tells a very clear and intuitive story: the CO$_2$ detection depends almost exclusively on the points in the CO$_2$ peak.  Coming up with a similarly simple and compelling story to explain which features of a transmission spectrum imply a certain metallicity and C/O ratio appears to be much more difficult.

\section{Conclusion}
In this paper, we presented a new and improved version of PLATON suitable for the JWST era.  We used this code to retrieve on the JWST/NIRCam 2.4--5.0 \um transmission and emission spectrum of HD 189733b, which we obtained from our own SPARTA reduction.  The retrievals indicate a metallicity of 2.6--6.8 times solar and a C/O around 0.42.  The transmission spectrum C/O posterior is broad, asymmetric, and 3$\sigma$ consistent with very low values ($<$0.2); the emission spectrum posterior is narrower and inconsistent with very low values.  We find strong evidence of H$_2$O, CO$_2$, CO, and H$_2$S in both transit and eclipse, but no evidence of CH$_4$; instead, we find that methane is depleted on the terminator from equilibrium abundances.  We find very tentative (1.8$\sigma$) evidence of a millibar thermal inversion on the dayside.  Better understanding of the mechanisms behind methane depletion (e.g. transport and photochemistry), of condensation, and of the correlated noise in NIRCam spectra is necessary for more precise constraints.

Retrievals on JWST-quality data are potentially affected by a large number of factors, including the cloud parameterization, non-equilibrium chemistry, the definition of metallicity and C/O, line list choice, and opacity resolution.  We explored some of these factors in the present paper using a single retrieval code, PLATON, but a comprehensive exploration of the robustness of retrieval inferences, while badly needed, is beyond our scope.  Such an exploration is being carried out by the Early Release Science team on all the JWST transmission spectra obtained so far for WASP-39b, spanning from 0.52 \um to 12 \um (Welbanks et al, in prep.)  We eagerly await their results.

\section{Acknowledgments}
We thank Heather Knutson and Eliza Kempton for valuable discussions, and for their contributions to earlier versions of PLATON. 
 MZ thanks the Heising-Simons Foundation for funding his 51 Pegasi b fellowship.  Computation for this project was performed on University of Chicago's Midway 3 cluster and Caltech's High Performance Computing (HPC) cluster.

\appendix
\section{Corner plots}
\begin{figure}
    \centering
    \includegraphics[width=\linewidth]{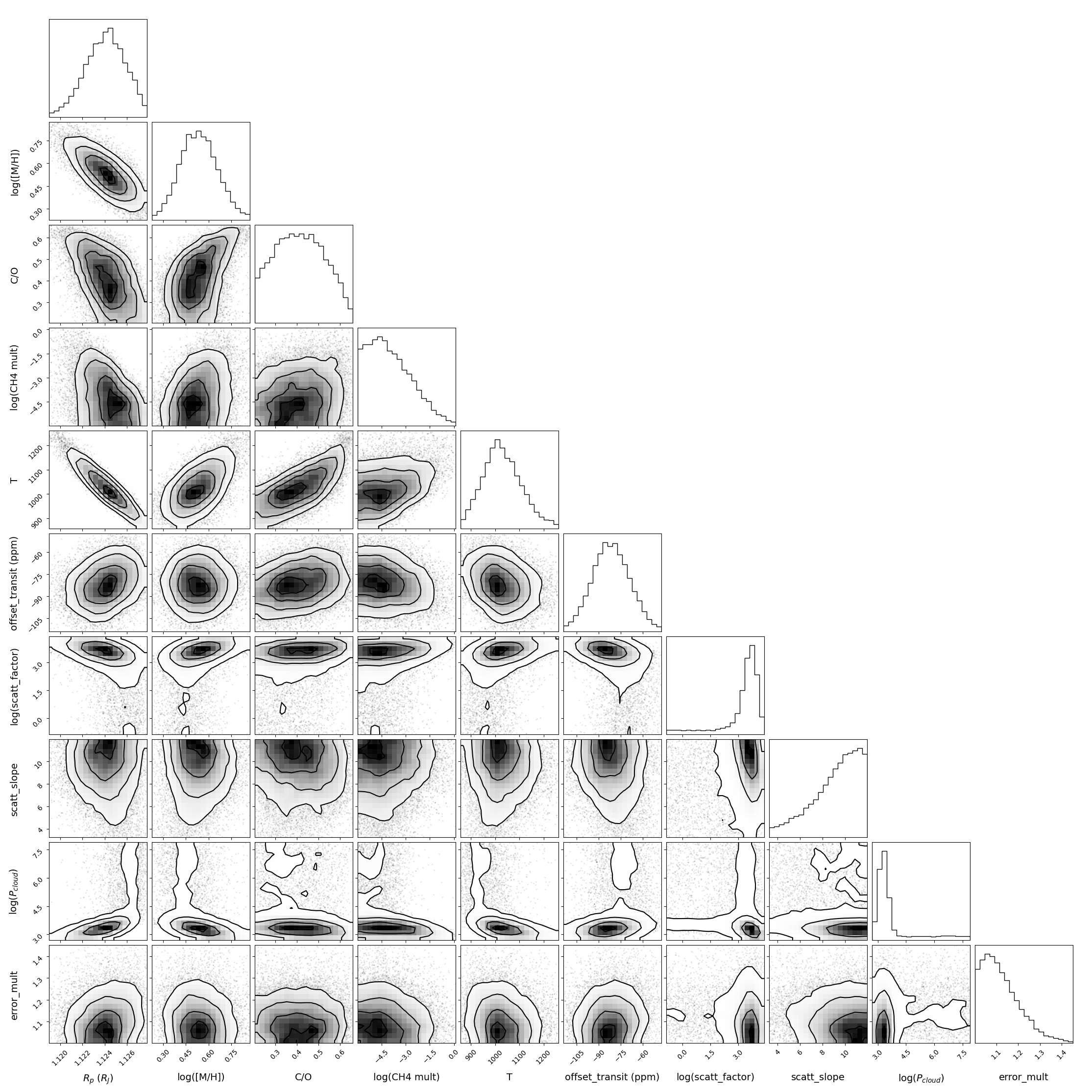}
    \caption{Posteriors for the fiducial NIRCam transmission spectrum retrieval, which assumes equilibrium chemistry with methane depletion.}
    \label{fig:transit_corner}
\end{figure}

\begin{figure}
    \centering
    \includegraphics[width=\linewidth]{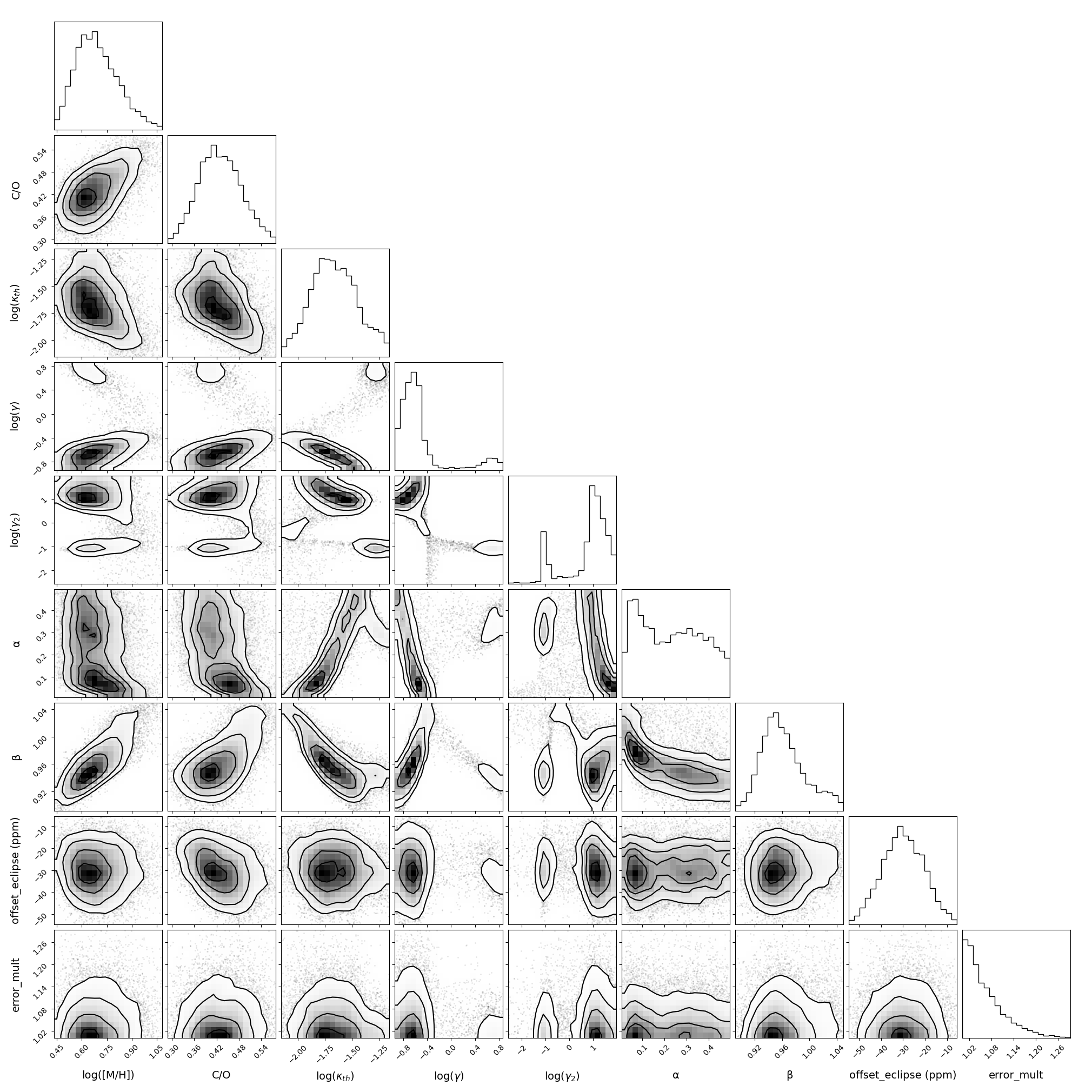}
    \caption{Posteriors for the fiducial NIRCam emission spectrum retrieval, which assumes equilibrium chemistry without methane depletion.}
    \label{fig:eclipse_corner}
\end{figure}

Figures \ref{fig:transit_corner} and \ref{fig:eclipse_corner} show the 2D posteriors (``corner plots'') for the fiducial NIRCam transit and eclipse retrievals.

\bibliographystyle{aasjournal} \bibliography{main}

\begin{thebibliography}{}
\expandafter\ifx\csname natexlab\endcsname\relax\def\natexlab#1{#1}\fi
\providecommand{\url}[1]{\href{#1}{#1}}

\bibitem[{{Addison} {et~al.}(2019){Addison}, {Wright}, {Wittenmyer}, {Horner}, {Mengel}, {Johns}, {Marti}, {Nicholson}, {Soutter}, {Bowler}, {Crossfield}, {Kane}, {Kielkopf}, {Plavchan}, {Tinney}, {Zhang}, {Clark}, {Clerte}, {Eastman}, {Swift}, {Bottom}, {Muirhead}, {McCrady}, {Herzig}, {Hogstrom}, {Wilson}, {Sliski}, {Johnson}, {Wright}, {Johnson}, {Blake}, {Riddle}, {Lin}, {Cornachione}, {Bedding}, {Stello}, {Huber}, {Marsden}, \& {Carter}}]{addison_2019}
{Addison}, B., {Wright}, D.~J., {Wittenmyer}, R.~A., {et~al.} 2019, \pasp, 131, 115003

\bibitem[{{Agol} {et~al.}(2010){Agol}, {Cowan}, {Knutson}, {Deming}, {Steffen}, {Henry}, \& {Charbonneau}}]{agol_2010}
{Agol}, E., {Cowan}, N.~B., {Knutson}, H.~A., {et~al.} 2010, \apj, 721, 1861

\bibitem[{Aitchison(1982)}]{aitchison_1982}
Aitchison, J. 1982, Journal of the Royal Statistical Society. Series B (Methodological), 44, 139.
\newblock \url{http://www.jstor.org/stable/2345821}

\bibitem[{{Allard} {et~al.}(2019){Allard}, {Spiegelman}, {Leininger}, \& {Molliere}}]{allard_2019}
{Allard}, N.~F., {Spiegelman}, F., {Leininger}, T., \& {Molliere}, P. 2019, \aap, 628, A120

\bibitem[{{Allard, N. F.} {et~al.}(2016){Allard, N. F.}, {Spiegelman, F.}, \& {Kielkopf, J. F.}}]{allard_2016}
{Allard, N. F.}, {Spiegelman, F.}, \& {Kielkopf, J. F.} 2016, A\&A, 589, A21.
\newblock \url{https://doi.org/10.1051/0004-6361/201628270}

\bibitem[{{Asplund} {et~al.}(2021){Asplund}, {Amarsi}, \& {Grevesse}}]{asplund_2021}
{Asplund}, M., {Amarsi}, A.~M., \& {Grevesse}, N. 2021, \aap, 653, A141

\bibitem[{{Asplund} {et~al.}(2009){Asplund}, {Grevesse}, {Sauval}, \& {Scott}}]{asplund_2009}
{Asplund}, M., {Grevesse}, N., {Sauval}, A.~J., \& {Scott}, P. 2009, \araa, 47, 481

\bibitem[{August {et~al.}(2023)August, Bean, Zhang, Lunine, Xue, Line, \& Smith}]{august_2023}
August, P.~C., Bean, J.~L., Zhang, M., {et~al.} 2023, The Astrophysical Journal Letters, 953, L24.
\newblock \url{https://dx.doi.org/10.3847/2041-8213/ace828}

\bibitem[{{Baxter} {et~al.}(2020){Baxter}, {D{\'e}sert}, {Parmentier}, {Line}, {Fortney}, {Arcangeli}, {Bean}, {Todorov}, \& {Mansfield}}]{baxter_2020}
{Baxter}, C., {D{\'e}sert}, J.-M., {Parmentier}, V., {et~al.} 2020, \aap, 639, A36

\bibitem[{{Bean} {et~al.}(2023){Bean}, {Xue}, {August}, {Lunine}, {Zhang}, {Thorngren}, {Tsai}, {Stassun}, {Schlawin}, {Ahrer}, {Ih}, \& {Mansfield}}]{bean_2023}
{Bean}, J.~L., {Xue}, Q., {August}, P.~C., {et~al.} 2023, \nat, 618, 43

\bibitem[{{Beaulieu} {et~al.}(2008){Beaulieu}, {Carey}, {Ribas}, \& {Tinetti}}]{beaulieu_2008}
{Beaulieu}, J.~P., {Carey}, S., {Ribas}, I., \& {Tinetti}, G. 2008, \apj, 677, 1343

\bibitem[{{Bell} {et~al.}(2024){Bell}, {Crouzet}, {Cubillos}, {Kreidberg}, {Piette}, {Roman}, {Barstow}, {Blecic}, {Carone}, {Coulombe}, {Ducrot}, {Hammond}, {Mendon{\c{c}}a}, {Moses}, {Parmentier}, {Stevenson}, {Teinturier}, {Zhang}, {Batalha}, {Bean}, {Benneke}, {Charnay}, {Chubb}, {Demory}, {Gao}, {Lee}, {L{\'o}pez-Morales}, {Morello}, {Rauscher}, {Sing}, {Tan}, {Venot}, {Wakeford}, {Aggarwal}, {Ahrer}, {Alam}, {Baeyens}, {Barrado}, {Caceres}, {Carter}, {Casewell}, {Challener}, {Crossfield}, {Decin}, {D{\'e}sert}, {Dobbs-Dixon}, {Dyrek}, {Espinoza}, {Feinstein}, {Gibson}, {Harrington}, {Helling}, {Hu}, {Iro}, {Kempton}, {Kendrew}, {Komacek}, {Krick}, {Lagage}, {Leconte}, {Lendl}, {Lewis}, {Lothringer}, {Malsky}, {Mancini}, {Mansfield}, {Mayne}, {Evans-Soma}, {Molaverdikhani}, {Nikolov}, {Nixon}, {Palle}, {Petit dit de la Roche}, {Piaulet}, {Powell}, {Rackham}, {Schneider}, {Steinrueck}, {Taylor}, {Welbanks}, {Yurchenko}, {Zhang}, \& {Zieba}}]{bell_2024}
{Bell}, T.~J., {Crouzet}, N., {Cubillos}, P.~E., {et~al.} 2024, Nature Astronomy, arXiv:2401.13027

\bibitem[{{Benneke} \& {Seager}(2012)}]{benneke_2012}
{Benneke}, B., \& {Seager}, S. 2012, \apj, 753, 100

\bibitem[{{Bernath}(2020)}]{bernath_2020}
{Bernath}, P.~F. 2020, \jqsrt, 240, 106687

\bibitem[{Blain {et~al.}(2024)Blain, Sánchez-López, \& Mollière}]{blain_2024}
Blain, D., Sánchez-López, A., \& Mollière, P. 2024, The Astronomical Journal, 167, 179.
\newblock \url{https://dx.doi.org/10.3847/1538-3881/ad2c8b}

\bibitem[{{Boucher} {et~al.}(2021){Boucher}, {Darveau-Bernier}, {Pelletier}, {Lafreni{\`e}re}, {Artigau}, {Cook}, {Allart}, {Radica}, {Doyon}, {Benneke}, {Arnold}, {Bonfils}, {Bourrier}, {Cloutier}, {Gomes da Silva}, {Deibert}, {Delfosse}, {Donati}, {Ehrenreich}, {Figueira}, {Forveille}, {Fouqu{\'e}}, {Gagn{\'e}}, {Gaidos}, {H{\'e}brard}, {Jayawardhana}, {Klein}, {Lovis}, {Martins}, {Martioli}, {Moutou}, \& {Santos}}]{boucher_2021}
{Boucher}, A., {Darveau-Bernier}, A., {Pelletier}, S., {et~al.} 2021, \aj, 162, 233

\bibitem[{{Bouchy} {et~al.}(2005){Bouchy}, {Udry}, {Mayor}, {Moutou}, {Pont}, {Iribarne}, {da Silva}, {Ilovaisky}, {Queloz}, {Santos}, {S{\'e}gransan}, \& {Zucker}}]{bouchy_2005}
{Bouchy}, F., {Udry}, S., {Mayor}, M., {et~al.} 2005, \aap, 444, L15

\bibitem[{{Bourrier} {et~al.}(2020){Bourrier}, {Wheatley}, {Lecavelier des Etangs}, {King}, {Louden}, {Ehrenreich}, {Fares}, {Helling}, {Llama}, {Jardine}, \& {Vidotto}}]{bourrier_2020}
{Bourrier}, V., {Wheatley}, P.~J., {Lecavelier des Etangs}, A., {et~al.} 2020, \mnras, 493, 559

\bibitem[{Bowesman {et~al.}(2024)Bowesman, Qu, McKemmish, Yurchenko, \& Tennyson}]{bowesman_2024}
Bowesman, C.~A., Qu, Q., McKemmish, L.~K., Yurchenko, S.~N., \& Tennyson, J. 2024, Monthly Notices of the Royal Astronomical Society, 529, 1321.
\newblock \url{https://doi.org/10.1093/mnras/stae542}

\bibitem[{Changeat {et~al.}(2022)Changeat, Edwards, Al-Refaie, Tsiaras, Skinner, Cho, Yip, Anisman, Ikoma, Bieger, Venot, Shibata, Waldmann, \& Tinetti}]{changeat_2022}
Changeat, Q., Edwards, B., Al-Refaie, A.~F., {et~al.} 2022, The Astrophysical Journal Supplement Series, 260, 3.
\newblock \url{https://dx.doi.org/10.3847/1538-4365/ac5cc2}

\bibitem[{{Charbonneau} {et~al.}(2008){Charbonneau}, {Knutson}, {Barman}, {Allen}, {Mayor}, {Megeath}, {Queloz}, \& {Udry}}]{charbonneau_2008}
{Charbonneau}, D., {Knutson}, H.~A., {Barman}, T., {et~al.} 2008, \apj, 686, 1341

\bibitem[{{Chubb} {et~al.}(2020){Chubb}, {Tennyson}, \& {Yurchenko}}]{chubb_2020}
{Chubb}, K.~L., {Tennyson}, J., \& {Yurchenko}, S.~N. 2020, \mnras, 493, 1531

\bibitem[{{Crouzet} {et~al.}(2014){Crouzet}, {McCullough}, {Deming}, \& {Madhusudhan}}]{crouzet_2014}
{Crouzet}, N., {McCullough}, P.~R., {Deming}, D., \& {Madhusudhan}, N. 2014, \apj, 795, 166

\bibitem[{{Deming} {et~al.}(2023){Deming}, {Line}, {Knutson}, {Crossfield}, {Kempton}, {Komacek}, {Wallack}, \& {Fu}}]{deming_2023}
{Deming}, D., {Line}, M.~R., {Knutson}, H.~A., {et~al.} 2023, \aj, 165, 104

\bibitem[{{Deming} \& {Seager}(2017)}]{deming_seager_2017}
{Deming}, L.~D., \& {Seager}, S. 2017, Journal of Geophysical Research (Planets), 122, 53

\bibitem[{{D{\'e}sert} {et~al.}(2011){D{\'e}sert}, {Sing}, {Vidal-Madjar}, {H{\'e}brard}, {Ehrenreich}, {Lecavelier Des Etangs}, {Parmentier}, {Ferlet}, \& {Henry}}]{desert_2011}
{D{\'e}sert}, J.~M., {Sing}, D., {Vidal-Madjar}, A., {et~al.} 2011, \aap, 526, A12

\bibitem[{{Dyrek} {et~al.}(2024){Dyrek}, {Min}, {Decin}, {Bouwman}, {Crouzet}, {Molli{\`e}re}, {Lagage}, {Konings}, {Tremblin}, {G{\"u}del}, {Pye}, {Waters}, {Henning}, {Vandenbussche}, {Ardevol Martinez}, {Argyriou}, {Ducrot}, {Heinke}, {van Looveren}, {Absil}, {Barrado}, {Baudoz}, {Boccaletti}, {Cossou}, {Coulais}, {Edwards}, {Gastaud}, {Glasse}, {Glauser}, {Greene}, {Kendrew}, {Krause}, {Lahuis}, {Mueller}, {Olofsson}, {Patapis}, {Rouan}, {Royer}, {Scheithauer}, {Waldmann}, {Whiteford}, {Colina}, {van Dishoeck}, {{\"O}stlin}, {Ray}, \& {Wright}}]{dyrek_2024}
{Dyrek}, A., {Min}, M., {Decin}, L., {et~al.} 2024, \nat, 625, 51

\bibitem[{Désert {et~al.}(2009)Désert, des Etangs, Hébrard, Sing, Ehrenreich, Ferlet, \& Vidal-Madjar}]{desert_2009}
Désert, J.-M., des Etangs, A.~L., Hébrard, G., {et~al.} 2009, The Astrophysical Journal, 699, 478.
\newblock \url{https://dx.doi.org/10.1088/0004-637X/699/1/478}

\bibitem[{Ehrenreich {et~al.}(2007)Ehrenreich, Hébrard, des Etangs, Sing, Désert, Bouchy, Ferlet, \& Vidal-Madjar}]{ehrenreich_2007}
Ehrenreich, D., Hébrard, G., des Etangs, A.~L., {et~al.} 2007, The Astrophysical Journal, 668, L179.
\newblock \url{https://dx.doi.org/10.1086/522792}

\bibitem[{{Evans} {et~al.}(2013){Evans}, {Pont}, {Sing}, {Aigrain}, {Barstow}, {D{\'e}sert}, {Gibson}, {Heng}, {Knutson}, \& {Lecavelier des Etangs}}]{evans_2013}
{Evans}, T.~M., {Pont}, F., {Sing}, D.~K., {et~al.} 2013, \apjl, 772, L16

\bibitem[{{Feroz} {et~al.}(2009){Feroz}, {Hobson}, \& {Bridges}}]{feroz_2009}
{Feroz}, F., {Hobson}, M.~P., \& {Bridges}, M. 2009, \mnras, 398, 1601

\bibitem[{{Feroz} {et~al.}(2019){Feroz}, {Hobson}, {Cameron}, \& {Pettitt}}]{feroz_2019}
{Feroz}, F., {Hobson}, M.~P., {Cameron}, E., \& {Pettitt}, A.~N. 2019, The Open Journal of Astrophysics, 2, 10

\bibitem[{Finnerty {et~al.}(2024)Finnerty, Xuan, Xin, Liberman, Schofield, Fitzgerald, Agrawal, Baker, Bartos, Blake, Calvin, Cetre, Delorme, Doppmann, Echeverri, Hsu, Jovanovic, López, Martin, Mawet, Morris, Pezzato, Ruffio, Sappey, Skemer, Venenciano, Wallace, Wallack, Wang, \& Wang}]{finnerty_2024}
Finnerty, L., Xuan, J.~W., Xin, Y., {et~al.} 2024, The Astronomical Journal, 167, 43.
\newblock \url{https://dx.doi.org/10.3847/1538-3881/ad1180}

\bibitem[{Flowers {et~al.}(2019)Flowers, Brogi, Rauscher, Kempton, \& Chiavassa}]{flowers_2019}
Flowers, E., Brogi, M., Rauscher, E., Kempton, E. M.-R., \& Chiavassa, A. 2019, The Astronomical Journal, 157, 209.
\newblock \url{https://dx.doi.org/10.3847/1538-3881/ab164c}

\bibitem[{{Fu} {et~al.}(2022){Fu}, {Espinoza}, {Sing}, {Lothringer}, {Dos Santos}, {Rustamkulov}, {Deming}, {Kempton}, {Komacek}, {Knutson}, {Albert}, {Pontoppidan}, {Volk}, \& {Filippazzo}}]{fu_2022}
{Fu}, G., {Espinoza}, N., {Sing}, D.~K., {et~al.} 2022, \apjl, 940, L35

\bibitem[{Fu {et~al.}(2024)Fu, Welbanks, Deming, Inglis, Zhang, Lothringer, Ih, Moses, Schlawin, Knutson, Henry, Greene, Sing, Savel, Kempton, Louie, Line, \& Nixon}]{fu_2024}
Fu, G., Welbanks, L., Deming, D., {et~al.} 2024, Nature, doi:10.1038/s41586-024-07760-y.
\newblock \url{https://doi.org/10.1038/s41586-024-07760-y}

\bibitem[{{Gandhi} \& {Madhusudhan}(2019)}]{gandhi_2019}
{Gandhi}, S., \& {Madhusudhan}, N. 2019, \mnras, 485, 5817

\bibitem[{{Gibson} {et~al.}(2011){Gibson}, {Pont}, \& {Aigrain}}]{gibson_2011}
{Gibson}, N.~P., {Pont}, F., \& {Aigrain}, S. 2011, \mnras, 411, 2199

\bibitem[{{Gibson} {et~al.}(2012){Gibson}, {Aigrain}, {Pont}, {Sing}, {D{\'e}sert}, {Evans}, {Henry}, {Husnoo}, \& {Knutson}}]{gibson_2012}
{Gibson}, N.~P., {Aigrain}, S., {Pont}, F., {et~al.} 2012, \mnras, 422, 753

\bibitem[{{Gordon} {et~al.}(2022){Gordon}, {Rothman}, {Hargreaves}, {Hashemi}, {Karlovets}, {Skinner}, {Conway}, {Hill}, {Kochanov}, {Tan}, {Wcis{\l}o}, {Finenko}, {Nelson}, {Bernath}, {Birk}, {Boudon}, {Campargue}, {Chance}, {Coustenis}, {Drouin}, {Flaud}, {Gamache}, {Hodges}, {Jacquemart}, {Mlawer}, {Nikitin}, {Perevalov}, {Rotger}, {Tennyson}, {Toon}, {Tran}, {Tyuterev}, {Adkins}, {Baker}, {Barbe}, {Can{\`e}}, {Cs{\'a}sz{\'a}r}, {Dudaryonok}, {Egorov}, {Fleisher}, {Fleurbaey}, {Foltynowicz}, {Furtenbacher}, {Harrison}, {Hartmann}, {Horneman}, {Huang}, {Karman}, {Karns}, {Kassi}, {Kleiner}, {Kofman}, {Kwabia-Tchana}, {Lavrentieva}, {Lee}, {Long}, {Lukashevskaya}, {Lyulin}, {Makhnev}, {Matt}, {Massie}, {Melosso}, {Mikhailenko}, {Mondelain}, {M{\"u}ller}, {Naumenko}, {Perrin}, {Polyansky}, {Raddaoui}, {Raston}, {Reed}, {Rey}, {Richard}, {T{\'o}bi{\'a}s}, {Sadiek}, {Schwenke}, {Starikova}, {Sung}, {Tamassia}, {Tashkun}, {Vander Auwera}, {Vasilenko}, {Vigasin}, {Villanueva}, {Vispoel}, {Wagner}, {Yachmenev}, \&
  {Yurchenko}}]{gordon_2022}
{Gordon}, I.~E., {Rothman}, L.~S., {Hargreaves}, R.~J., {et~al.} 2022, \jqsrt, 277, 107949

\bibitem[{{Grimm} {et~al.}(2021){Grimm}, {Malik}, {Kitzmann}, {Guzm{\'a}n-Mesa}, {Hoeijmakers}, {Fisher}, {Mendon{\c{c}}a}, {Yurchenko}, {Tennyson}, {Alesina}, {Buchschacher}, {Burnier}, {Segransan}, {Kurucz}, \& {Heng}}]{grimm_2021}
{Grimm}, S.~L., {Malik}, M., {Kitzmann}, D., {et~al.} 2021, \apjs, 253, 30

\bibitem[{{Henry} \& {Winn}(2008)}]{henry_2008}
{Henry}, G.~W., \& {Winn}, J.~N. 2008, \aj, 135, 68

\bibitem[{Ih \& Kempton(2021)}]{ih_2021}
Ih, J., \& Kempton, E. M.-R. 2021, The Astronomical Journal, 162, 237.
\newblock \url{https://dx.doi.org/10.3847/1538-3881/ac173b}

\bibitem[{{Kempton} {et~al.}(2017){Kempton}, {Lupu}, {Owusu-Asare}, {Slough}, \& {Cale}}]{kempton_2017}
{Kempton}, E.~M.-R., {Lupu}, R., {Owusu-Asare}, A., {Slough}, P., \& {Cale}, B. 2017, \pasp, 129, 044402

\bibitem[{{Kempton} {et~al.}(2023){Kempton}, {Zhang}, {Bean}, {Steinrueck}, {Piette}, {Parmentier}, {Malsky}, {Roman}, {Rauscher}, {Gao}, {Bell}, {Xue}, {Taylor}, {Savel}, {Arnold}, {Nixon}, {Stevenson}, {Mansfield}, {Kendrew}, {Zieba}, {Ducrot}, {Dyrek}, {Lagage}, {Stassun}, {Henry}, {Barman}, {Lupu}, {Malik}, {Kataria}, {Ih}, {Fu}, {Welbanks}, \& {McGill}}]{kempton_2023}
{Kempton}, E. M.~R., {Zhang}, M., {Bean}, J.~L., {et~al.} 2023, \nat, 620, 67

\bibitem[{{Kilpatrick} {et~al.}(2020){Kilpatrick}, {Kataria}, {Lewis}, {Zellem}, {Henry}, {Cowan}, {de Wit}, {Fortney}, {Knutson}, {Seager}, {Showman}, \& {Tucker}}]{kilpatrick_2020}
{Kilpatrick}, B.~M., {Kataria}, T., {Lewis}, N.~K., {et~al.} 2020, \aj, 159, 51

\bibitem[{King(2019)}]{king_2019}
King, G.~W. 2019, unpublished.
\newblock \url{http://webcat.warwick.ac.uk/record=b3490502~S15}

\bibitem[{{Kitzmann} \& {Heng}(2018)}]{kitzmann_2018}
{Kitzmann}, D., \& {Heng}, K. 2018, \mnras, 475, 94

\bibitem[{{Kitzmann} {et~al.}(2024){Kitzmann}, {Stock}, \& {Patzer}}]{kitzmann_2024}
{Kitzmann}, D., {Stock}, J.~W., \& {Patzer}, A. B.~C. 2024, \mnras, 527, 7263

\bibitem[{{Klein} {et~al.}(2024){Klein}, {Debras}, {Donati}, {Hood}, {Moutou}, {Carmona}, {Ould-elkhim}, {B{\'e}zard}, {Charnay}, {Fouqu{\'e}}, {Masson}, {Vinatier}, {Baruteau}, {Boisse}, {Bonfils}, {Chiavassa}, {Delfosse}, {Dethier}, {Hebrard}, {Kiefer}, {Leconte}, {Martioli}, {Parmentier}, {Petit}, {Pluriel}, {Selsis}, {Teinturier}, {Tremblin}, {Turbet}, {Venot}, \& {Wyttenbach}}]{klein_2024}
{Klein}, B., {Debras}, F., {Donati}, J.-F., {et~al.} 2024, \mnras, 527, 544

\bibitem[{{Knutson} {et~al.}(2007){Knutson}, {Charbonneau}, {Noyes}, {Brown}, \& {Gilliland}}]{knutson_2007}
{Knutson}, H.~A., {Charbonneau}, D., {Noyes}, R.~W., {Brown}, T.~M., \& {Gilliland}, R.~L. 2007, \apj, 655, 564

\bibitem[{{Knutson} {et~al.}(2009){Knutson}, {Charbonneau}, {Cowan}, {Fortney}, {Showman}, {Agol}, {Henry}, {Everett}, \& {Allen}}]{knutson_2009}
{Knutson}, H.~A., {Charbonneau}, D., {Cowan}, N.~B., {et~al.} 2009, \apj, 690, 822

\bibitem[{{Knutson} {et~al.}(2012){Knutson}, {Lewis}, {Fortney}, {Burrows}, {Showman}, {Cowan}, {Agol}, {Aigrain}, {Charbonneau}, \& {Deming}}]{knutson_2012}
{Knutson}, H.~A., {Lewis}, N., {Fortney}, J.~J., {et~al.} 2012, \apj, 754, 22

\bibitem[{{Kreidberg} {et~al.}(2018){Kreidberg}, {Line}, {Thorngren}, {Morley}, \& {Stevenson}}]{kreidberg_2018}
{Kreidberg}, L., {Line}, M.~R., {Thorngren}, D., {Morley}, C.~V., \& {Stevenson}, K.~B. 2018, \apjl, 858, L6

\bibitem[{{Lamp{\'o}n} {et~al.}(2021){Lamp{\'o}n}, {L{\'o}pez-Puertas}, {Sanz-Forcada}, {S{\'a}nchez-L{\'o}pez}, {Molaverdikhani}, {Czesla}, {Quirrenbach}, {Pall{\'e}}, {Caballero}, {Henning}, {Salz}, {Nortmann}, {Aceituno}, {Amado}, {Bauer}, {Montes}, {Nagel}, {Reiners}, \& {Ribas}}]{lampon_2021}
{Lamp{\'o}n}, M., {L{\'o}pez-Puertas}, M., {Sanz-Forcada}, J., {et~al.} 2021, \aap, 647, A129

\bibitem[{{Lee} {et~al.}(2015){Lee}, {Helling}, {Dobbs-Dixon}, \& {Juncher}}]{lee_2015}
{Lee}, G., {Helling}, C., {Dobbs-Dixon}, I., \& {Juncher}, D. 2015, \aap, 580, A12

\bibitem[{{Lee} {et~al.}(2012){Lee}, {Fletcher}, \& {Irwin}}]{lee_2012}
{Lee}, J.~M., {Fletcher}, L.~N., \& {Irwin}, P.~G.~J. 2012, \mnras, 420, 170

\bibitem[{Lee {et~al.}(2014)Lee, Irwin, Fletcher, Heng, \& Barstow}]{lee_2014}
Lee, J.-M., Irwin, P. G.~J., Fletcher, L.~N., Heng, K., \& Barstow, J.~K. 2014, The Astrophysical Journal, 789, 14.
\newblock \url{https://doi.org/10.1088%2F0004-637x%2F789%2F1%2F14}

\bibitem[{{Line} {et~al.}(2013){Line}, {Wolf}, {Zhang}, {Knutson}, {Kammer}, {Ellison}, {Deroo}, {Crisp}, \& {Yung}}]{line_2013}
{Line}, M.~R., {Wolf}, A.~S., {Zhang}, X., {et~al.} 2013, \apj, 775, 137

\bibitem[{{Lines} {et~al.}(2018){Lines}, {Mayne}, {Boutle}, {Manners}, {Lee}, {Helling}, {Drummond}, {Amundsen}, {Goyal}, {Acreman}, {Tremblin}, \& {Kerslake}}]{lines_2018}
{Lines}, S., {Mayne}, N.~J., {Boutle}, I.~A., {et~al.} 2018, \aap, 615, A97

\bibitem[{{Lodders} {et~al.}(2009){Lodders}, {Palme}, \& {Gail}}]{lodders_2009}
{Lodders}, K., {Palme}, H., \& {Gail}, H.~P. 2009, Landolt B\&ouml;rnstein, 4B, 712

\bibitem[{Magic {et~al.}(2015)Magic, Chiavassa, Collet, \& Asplund}]{magic_2015}
Magic, Z., Chiavassa, A., Collet, R., \& Asplund, M. 2015, Astronomy \& Astrophysics, 573, A90

\bibitem[{{Mansfield} {et~al.}(2021){Mansfield}, {Line}, {Bean}, {Fortney}, {Parmentier}, {Wiser}, {Kempton}, {Gharib-Nezhad}, {Sing}, {L{\'o}pez-Morales}, {Baxter}, {D{\'e}sert}, {Swain}, \& {Roudier}}]{mansfield_2021}
{Mansfield}, M., {Line}, M.~R., {Bean}, J.~L., {et~al.} 2021, Nature Astronomy, 5, 1224

\bibitem[{{McCullough} {et~al.}(2014){McCullough}, {Crouzet}, {Deming}, \& {Madhusudhan}}]{mccullough_2014}
{McCullough}, P.~R., {Crouzet}, N., {Deming}, D., \& {Madhusudhan}, N. 2014, \apj, 791, 55

\bibitem[{{Morello} {et~al.}(2023){Morello}, {Changeat}, {Dyrek}, {Lagage}, \& {Tan}}]{morello_2023}
{Morello}, G., {Changeat}, Q., {Dyrek}, A., {Lagage}, P.~O., \& {Tan}, J.~C. 2023, \aap, 676, A54

\bibitem[{{Morello} {et~al.}(2014){Morello}, {Waldmann}, {Tinetti}, {Peres}, {Micela}, \& {Howarth}}]{morello_2014}
{Morello}, G., {Waldmann}, I.~P., {Tinetti}, G., {et~al.} 2014, \apj, 786, 22

\bibitem[{{Moses} {et~al.}(2013){Moses}, {Madhusudhan}, {Visscher}, \& {Freedman}}]{moses_2013}
{Moses}, J.~I., {Madhusudhan}, N., {Visscher}, C., \& {Freedman}, R.~S. 2013, \apj, 763, 25

\bibitem[{{Moses} {et~al.}(2011){Moses}, {Visscher}, {Fortney}, {Showman}, {Lewis}, {Griffith}, {Klippenstein}, {Shabram}, {Friedson}, \& {Marley}}]{moses_2011}
{Moses}, J.~I., {Visscher}, C., {Fortney}, J.~J., {et~al.} 2011, \apj, 737, 15

\bibitem[{{{\"O}berg} {et~al.}(2011){{\"O}berg}, {Murray-Clay}, \& {Bergin}}]{oberg_2011}
{{\"O}berg}, K.~I., {Murray-Clay}, R., \& {Bergin}, E.~A. 2011, \apjl, 743, L16

\bibitem[{{Ohno} \& {Kawashima}(2020)}]{ohno_2020}
{Ohno}, K., \& {Kawashima}, Y. 2020, \apjl, 895, L47

\bibitem[{Owens {et~al.}(2024)Owens, Yurchenko, \& Tennyson}]{owens_2024}
Owens, A., Yurchenko, S.~N., \& Tennyson, J. 2024, Monthly Notices of the Royal Astronomical Society, 530, 4004.
\newblock \url{https://doi.org/10.1093/mnras/stae1110}

\bibitem[{{Pinhas} {et~al.}(2019){Pinhas}, {Madhusudhan}, {Gandhi}, \& {MacDonald}}]{pinhas_2019}
{Pinhas}, A., {Madhusudhan}, N., {Gandhi}, S., \& {MacDonald}, R. 2019, \mnras, 482, 1485

\bibitem[{{Plazas} {et~al.}(2018){Plazas}, {Shapiro}, {Smith}, {Huff}, \& {Rhodes}}]{plazas_2018}
{Plazas}, A.~A., {Shapiro}, C., {Smith}, R., {Huff}, E., \& {Rhodes}, J. 2018, \pasp, 130, 065004

\bibitem[{{Pont} {et~al.}(2013){Pont}, {Sing}, {Gibson}, {Aigrain}, {Henry}, \& {Husnoo}}]{pont_2013}
{Pont}, F., {Sing}, D.~K., {Gibson}, N.~P., {et~al.} 2013, \mnras, 432, 2917

\bibitem[{{Poppenhaeger} {et~al.}(2013){Poppenhaeger}, {Schmitt}, \& {Wolk}}]{poppenhaeger_2013}
{Poppenhaeger}, K., {Schmitt}, J.~H.~M.~M., \& {Wolk}, S.~J. 2013, \apj, 773, 62

\bibitem[{Qu {et~al.}(2021)Qu, Yurchenko, \& Tennyson}]{qu_2021}
Qu, Q., Yurchenko, S.~N., \& Tennyson, J. 2021, Monthly Notices of the Royal Astronomical Society, 504, 5768.
\newblock \url{https://doi.org/10.1093/mnras/stab1154}

\bibitem[{{Rustamkulov} {et~al.}(2023){Rustamkulov}, {Sing}, {Mukherjee}, {May}, {Kirk}, {Schlawin}, {Line}, {Piaulet}, {Carter}, {Batalha}, {Goyal}, {L{\'o}pez-Morales}, {Lothringer}, {MacDonald}, {Moran}, {Stevenson}, {Wakeford}, {Espinoza}, {Bean}, {Batalha}, {Benneke}, {Berta-Thompson}, {Crossfield}, {Gao}, {Kreidberg}, {Powell}, {Cubillos}, {Gibson}, {Leconte}, {Molaverdikhani}, {Nikolov}, {Parmentier}, {Roy}, {Taylor}, {Turner}, {Wheatley}, {Aggarwal}, {Ahrer}, {Alam}, {Alderson}, {Allen}, {Banerjee}, {Barat}, {Barrado}, {Barstow}, {Bell}, {Blecic}, {Brande}, {Casewell}, {Changeat}, {Chubb}, {Crouzet}, {Daylan}, {Decin}, {D{\'e}sert}, {Mikal-Evans}, {Feinstein}, {Flagg}, {Fortney}, {Harrington}, {Heng}, {Hong}, {Hu}, {Iro}, {Kataria}, {Kempton}, {Krick}, {Lendl}, {Lillo-Box}, {Louca}, {Lustig-Yaeger}, {Mancini}, {Mansfield}, {Mayne}, {Miguel}, {Morello}, {Ohno}, {Palle}, {Petit dit de la Roche}, {Rackham}, {Radica}, {Ramos-Rosado}, {Redfield}, {Rogers}, {Shkolnik}, {Southworth}, {Teske}, {Tremblin},
  {Tucker}, {Venot}, {Waalkes}, {Welbanks}, {Zhang}, \& {Zieba}}]{rumstamkulov_2023}
{Rustamkulov}, Z., {Sing}, D.~K., {Mukherjee}, S., {et~al.} 2023, \nat, 614, 659

\bibitem[{{Sing} {et~al.}(2008){Sing}, {Vidal-Madjar}, {Lecavelier des Etangs}, {D{\'e}sert}, {Ballester}, \& {Ehrenreich}}]{sing_2008}
{Sing}, D.~K., {Vidal-Madjar}, A., {Lecavelier des Etangs}, A., {et~al.} 2008, \apj, 686, 667

\bibitem[{{Sing} {et~al.}(2011){Sing}, {Pont}, {Aigrain}, {Charbonneau}, {D{\'e}sert}, {Gibson}, {Gilliland}, {Hayek}, {Henry}, {Knutson}, {Lecavelier Des Etangs}, {Mazeh}, \& {Shporer}}]{sing_2011}
{Sing}, D.~K., {Pont}, F., {Aigrain}, S., {et~al.} 2011, \mnras, 416, 1443

\bibitem[{Speagle(2020)}]{speagle_2019}
Speagle, J.~S. 2020, Monthly Notices of the Royal Astronomical Society, 493, 3132.
\newblock \url{https://doi.org/10.1093/mnras/staa278}

\bibitem[{{Stevenson} {et~al.}(2010){Stevenson}, {Harrington}, {Nymeyer}, {Madhusudhan}, {Seager}, {Bowman}, {Hardy}, {Deming}, {Rauscher}, \& {Lust}}]{stevenson_2010}
{Stevenson}, K.~B., {Harrington}, J., {Nymeyer}, S., {et~al.} 2010, \nat, 464, 1161

\bibitem[{{Tinetti} {et~al.}(2007){Tinetti}, {Vidal-Madjar}, {Liang}, {Beaulieu}, {Yung}, {Carey}, {Barber}, {Tennyson}, {Ribas}, {Allard}, {Ballester}, {Sing}, \& {Selsis}}]{tinetti_2007}
{Tinetti}, G., {Vidal-Madjar}, A., {Liang}, M.-C., {et~al.} 2007, \nat, 448, 169

\bibitem[{Tsai {et~al.}(2017)Tsai, Lyons, Grosheintz, Rimmer, Kitzmann, \& Heng}]{tsai_2017}
Tsai, S.-M., Lyons, J.~R., Grosheintz, L., {et~al.} 2017, The Astrophysical Journal Supplement Series, 228, 20.
\newblock \url{https://dx.doi.org/10.3847/1538-4365/228/2/20}

\bibitem[{{Tsai} {et~al.}(2021){Tsai}, {Malik}, {Kitzmann}, {Lyons}, {Fateev}, {Lee}, \& {Heng}}]{tsai_2021}
{Tsai}, S.-M., {Malik}, M., {Kitzmann}, D., {et~al.} 2021, \apj, 923, 264

\bibitem[{{Tsai} {et~al.}(2023){Tsai}, {Lee}, {Powell}, {Gao}, {Zhang}, {Moses}, {H{\'e}brard}, {Venot}, {Parmentier}, {Jordan}, {Hu}, {Alam}, {Alderson}, {Batalha}, {Bean}, {Benneke}, {Bierson}, {Brady}, {Carone}, {Carter}, {Chubb}, {Inglis}, {Leconte}, {Line}, {L{\'o}pez-Morales}, {Miguel}, {Molaverdikhani}, {Rustamkulov}, {Sing}, {Stevenson}, {Wakeford}, {Yang}, {Aggarwal}, {Baeyens}, {Barat}, {de Val-Borro}, {Daylan}, {Fortney}, {France}, {Goyal}, {Grant}, {Kirk}, {Kreidberg}, {Louca}, {Moran}, {Mukherjee}, {Nasedkin}, {Ohno}, {Rackham}, {Redfield}, {Taylor}, {Tremblin}, {Visscher}, {Wallack}, {Welbanks}, {Youngblood}, {Ahrer}, {Batalha}, {Behr}, {Berta-Thompson}, {Blecic}, {Casewell}, {Crossfield}, {Crouzet}, {Cubillos}, {Decin}, {D{\'e}sert}, {Feinstein}, {Gibson}, {Harrington}, {Heng}, {Henning}, {Kempton}, {Krick}, {Lagage}, {Lendl}, {Lothringer}, {Mansfield}, {Mayne}, {Mikal-Evans}, {Palle}, {Schlawin}, {Shorttle}, {Wheatley}, \& {Yurchenko}}]{tsai_2023}
{Tsai}, S.-M., {Lee}, E. K.~H., {Powell}, D., {et~al.} 2023, \nat, 617, 483

\bibitem[{{Vehtari} {et~al.}(2015){Vehtari}, {Gelman}, \& {Gabry}}]{vehtari_2015}
{Vehtari}, A., {Gelman}, A., \& {Gabry}, J. 2015, arXiv e-prints, arXiv:1507.04544

\bibitem[{Vehtari \& Ojanen(2012)}]{vehtari_2012}
Vehtari, A., \& Ojanen, J. 2012, STATISTICS SURVEYS, 142

\bibitem[{{Welbanks} {et~al.}(2023){Welbanks}, {McGill}, {Line}, \& {Madhusudhan}}]{welbanks_2023}
{Welbanks}, L., {McGill}, P., {Line}, M., \& {Madhusudhan}, N. 2023, \aj, 165, 112

\bibitem[{{Welbanks} {et~al.}(2024){Welbanks}, {Bell}, {Beatty}, {Line}, {Ohno}, {Fortney}, {Schlawin}, {Greene}, {Rauscher}, {McGill}, {Murphy}, {Parmentier}, {Tang}, {Edelman}, {Mukherjee}, {Wiser}, {Lagage}, {Dyrek}, \& {Arnold}}]{welbanks_2024}
{Welbanks}, L., {Bell}, T.~J., {Beatty}, T.~G., {et~al.} 2024, arXiv e-prints, arXiv:2405.11018

\bibitem[{{Woitke} {et~al.}(2018){Woitke}, {Helling}, {Hunter}, {Millard}, {Turner}, {Worters}, {Blecic}, \& {Stock}}]{woitke_2018}
{Woitke}, P., {Helling}, C., {Hunter}, G.~H., {et~al.} 2018, \aap, 614, A1

\bibitem[{{Xue} {et~al.}(2024){Xue}, {Bean}, {Zhang}, {Welbanks}, {Lunine}, \& {August}}]{xue_2024}
{Xue}, Q., {Bean}, J.~L., {Zhang}, M., {et~al.} 2024, \apjl, 963, L5

\bibitem[{{Yurchenko} {et~al.}(2018){Yurchenko}, {Al-Refaie}, \& {Tennyson}}]{yurchenko_2018}
{Yurchenko}, S.~N., {Al-Refaie}, A.~F., \& {Tennyson}, J. 2018, \aap, 614, A131

\bibitem[{{Yurchenko} {et~al.}(2020){Yurchenko}, {Mellor}, {Freedman}, \& {Tennyson}}]{yurchenko_2020}
{Yurchenko}, S.~N., {Mellor}, T.~M., {Freedman}, R.~S., \& {Tennyson}, J. 2020, \mnras, 496, 5282

\bibitem[{Yurchenko {et~al.}(2024)Yurchenko, Owens, Kefala, \& Tennyson}]{yurchenko_2024}
Yurchenko, S.~N., Owens, A., Kefala, K., \& Tennyson, J. 2024, Monthly Notices of the Royal Astronomical Society, 528, 3719.
\newblock \url{https://doi.org/10.1093/mnras/stae148}

\bibitem[{{Yurchenko} {et~al.}(2022){Yurchenko}, {Tennyson}, {Syme}, {Adam}, {Clark}, {Cooper}, {Dobney}, {Donnelly}, {Gorman}, {Lynas-Gray}, {Meltzer}, {Owens}, {Qu}, {Semenov}, {Somogyi}, {Upadhyay}, {Wright}, \& {Zapata Trujillo}}]{yurchenko_2022}
{Yurchenko}, S.~N., {Tennyson}, J., {Syme}, A.-M., {et~al.} 2022, \mnras, 510, 903

\bibitem[{{Zhang} {et~al.}(2020){Zhang}, {Chachan}, {Kempton}, {Knutson}, \& {Chang}}]{zhang_2020}
{Zhang}, M., {Chachan}, Y., {Kempton}, E. M.~R., {Knutson}, H.~A., \& {Chang}, W.~H. 2020, \apj, 899, 27

\bibitem[{{Zhang} {et~al.}(2018){Zhang}, {Knutson}, {Kataria}, {Schwartz}, {Cowan}, {Showman}, {Burrows}, {Fortney}, {Todorov}, \& {Desert}}]{zhang_2017}
{Zhang}, M., {Knutson}, H.~A., {Kataria}, T., {et~al.} 2018, \aj, 155, 83

\end{thebibliography}
\end{document}